\documentclass[aps,prfluids,showpacs,amsfonts,notitlepage,amssymb,amsmath,nofootinbib,superscriptaddress]{revtex4-2}
\usepackage[utf8]{inputenc}
\usepackage[english]{babel}
\usepackage[titletoc,toc,title]{appendix}
\usepackage{amsmath}
\usepackage{amsfonts}
\usepackage{amssymb}
\usepackage[colorlinks=true,linkcolor=blue,citecolor=blue,urlcolor=blue]{hyperref}
\usepackage{graphicx}
\usepackage{enumerate}
\usepackage{csquotes}
\usepackage[caption=false]{subfig}
\usepackage{dsfont}
\usepackage{color}
\usepackage{bbold}
\usepackage{mathtools}
\usepackage{tensor}
\usepackage{xcolor}
\usepackage[normalem]{ulem}

\begin{document}

\title{Black hole analogues in two-dimensional flows with constant shear}

\author{Alessia Biondi}
\affiliation{Institut Pprime, CNRS--Université de Poitiers--ISAE-ENSMA. TSA 51124, 86073 Poitiers Cedex 9, France}
\author{Scott Robertson}
\affiliation{Institut Pprime, CNRS--Université de Poitiers--ISAE-ENSMA. TSA 51124, 86073 Poitiers Cedex 9, France}
\author{Germain Rousseaux}
\affiliation{Institut Pprime, CNRS--Université de Poitiers--ISAE-ENSMA. TSA 51124, 86073 Poitiers Cedex 9, France}

\date{\today}

\begin{abstract}
    We investigate the gravitational analogy for shallow water waves propagating on a unidirectional  
    background flow with constant shear. Generalizing the standard irrotational treatment, we demonstrate that a transverse background vorticity is perfectly compatible with the existence of an effective curved spacetime for longitudinal surface waves. We extract the modified acoustic metric and show that the conformal factor, which governs wave scattering, becomes strongly dependent on the vorticity. Consequently, we find that the presence of background shear significantly suppresses the scattering of modes over an inhomogeneous bathymetry, while mildly modifying the analogue surface gravity at the transcritical horizon. These results highlight the nontrivial role of rotational dynamics in analogue gravity experiments.
\end{abstract}

\maketitle

%%%%%%%%%%%%%%%%%%%%%%%%%%%%%%%%%%%%%%%%%%%%%%%%%%%%%%%%%%%%%%%%%%%%%%%%%%%%%%%%%%%%%%%%%%%%%%%%%%%%%%%%%%%%%%%%%%%%%%%%%

\section{Introduction}

Analogue Gravity~\cite{barcelo2026analogue} is a transdisciplinary field that aims to mimic field propagation in curved spacetime using waves propagating in a laboratory-based system. It exploits the mathematical equivalence between the kinematic equations of fields propagating on a curved spacetime, typically the Laplace-Beltrami equation for massless scalar fields
\begin{equation}
    \frac{1}{\sqrt{-g}} \partial_{\mu} \left( \sqrt{-g} \, g^{\mu\nu} \partial_{\nu} \phi \right) = 0 \,,
    \label{eq:intro_KG}
\end{equation}
and the propagation of excitations in inhomogeneous condensed matter or fluid systems through the identification of an {\it effective metric} that allows the wave equation to be written in the form~(\ref{eq:intro_KG}).  Unruh was the first to suggest a concrete use of the analogy~\cite{PhysRevLett.46.1351} by pointing out the possibility of analogue Hawking radiation from horizons in the effective spacetime.  In the past two decades, Analogue Gravity has become an experimental field, with realizations in systems such as Bose-Einstein condensates~\cite{garay2000sonic,CarlosBarceló_2001,PhysRevLett.105.240401,PhysRevLett.108.071101,deNova-2019}, optical fibers and waveguides~\cite{doi:10.1126/science.1153625,PhysRevLett.122.010404}, polariton fluids~\cite{PhysRevB.86.144505,jacquet_polariton_2020,PhysRevD.109.105024}, and surface waves on water~\cite{PhysRevD.66.044019,Weinfurtner_2011,Euve_2016,torres_rotational_2017,Euv__2020}.

The surface wave analogy was first expounded by Schützhold and Unruh~\cite{PhysRevD.66.044019}.  It yields an effective metric that, typically for waves in moving media, takes the form of a generalized Painlev\'{e}-Gullstrand metric
\begin{equation}
    ds^2 = g_{\mu\nu}dx^{\mu}dx^{\nu} = \Lambda^2(x) \left[ - \left( c^2(x) - U^2(x) \right) dt^2 - 2 U(x) \, dt \, dx + dx^2 + dy^2 \right] \,,
    \label{eq:intro_metric}
\end{equation}
where $U(x)$ is the generic background flow velocity, $c(x)$ is the local wave speed, and $\Lambda(x)$ is a conformal factor.  Intuitively, this yields a horizon wherever the flow velocity $U$ crosses the wave speed $c$, causing the $g_{tt}$ component of the effective metric to vanish.
This analogy has led to several experimental studies, including the observation of an analogue of the Hawking effect in subcritical flows~\cite{Weinfurtner_2011,Euve_2016}, rotational super-radiance at the ergoregion of a rotating flow~\cite{torres_rotational_2017}, and scattering of an infalling wave at a black-hole horizon~\cite{Euv__2020}.  The latter phenomenon is a result of wave scattering due to inhomogeneities in the conformal factor $\Lambda(x)$, which through an appropriate change of coordinates can be written as an effective potential barrier~\cite{Anderson2014} in a Regge-Wheeler-like equation~\cite{PhysRev.108.1063}, and yields an effective greybody factor~\cite{PhysRevD.14.3260} when considering the transmission of outgoing waves from a black-hole horizon.   The water wave system is thus very relevant to Analogue Gravity, despite its classical, macroscopic nature, and each of the parameters entering the effective metric~(\ref{eq:intro_metric}) has physical significance that maps onto its astrophysical analogue.  

However, real channel flows tend to be far more complicated in practice than simple theoretical descriptions would suggest, being subject to vorticity, viscosity, friction, nonlinearities, and turbulence.  All of these may be expected to distort the wave behavior away from the gravitational analogy, potentially rendering the system less relevant.  It thus behooves us to investigate these features and to assess whether they might be incorporated into an Analogue Gravity framework.

In this paper, we focus on vorticity, and in particular on the simple case of a two-dimensional flow with constant shear (so that the flow velocity varies essentially linearly with depth).  We generalise the Analogue Gravity framework to longitudinal surface waves on a constant-shear flow, showing that these waves continue to obey Eq.~(\ref{eq:intro_KG}) for an effective metric with the exact structural form of Eq.~(\ref{eq:intro_metric}), but with metric parameters that acquire a non-trivial dependence on the constant transverse shear vorticity $\Omega$.
We show that the vorticity fundamentally alters both the kinematic parameters and the conformal factor entering the effective metric, which has immediate consequences for wave scattering over an inhomogeneous bathymetry.
In particular, we show that the Hawking effect is mildly enhanced by the presence of shear, while the wave scattering responsible for the greybody factor is suppressed by it.
Note that, while previous studies within Analogue Gravity have examined the effects of vorticity on analogue rotating black holes (typically with vorticity oriented perpendicular to the fluid plane~\cite{PhysRevD.99.044025,PhysRevLett.88.110201}), the impact of a transverse vorticity due to shear on the longitudinal propagation of surface waves, and its significance for the Hawking effect, remains largely unexplored. Conversely, while the fluid mechanics literature has extensively derived the dispersion relations for waves on constant-shear flows~\cite{nepf1994,maïssa2013influenceshearflowvorticitywavecurrent,ellingsen2014linear,maissa2016negative,maissa2016wave}, these studies do not address the framework of an effective spacetime nor the consequences for wave  
scattering.

It is necessary to motivate our specific choice of velocity profile. In any experimental flume featuring a bottom obstacle, friction naturally generates a velocity gradient and a boundary layer. While realistic turbulent flows will tend to be more complicated and to have a vorticity profile that evolves due to viscosity, we  
focus here on an idealized 
model with zero viscosity and constant vorticity, 
for two critical reasons. First, this model guarantees analytical tractability. We are fundamentally interested in deriving an explicit equation of motion that incorporates a spatially varying bottom $b(x)$, which is strictly required to capture the analogue horizon and treat wave scattering analytically. Considering more complex velocity profiles over a varying bottom rapidly becomes analytically intractable, obscuring the physical mechanisms of the analogue Hawking effect. Second, this linear shear profile is not merely a mathematical convenience; it is experimentally realizable. It has been 
demonstrated in the fluid mechanics literature that laboratories can successfully generate and sustain nearly-constant-shear flows in water channels~\cite{nepf1994,dunn2007}.  Therefore, our constant-shear setup represents the optimal balance: it cleanly isolates the  
rotational dynamics of the fluid while preserving the exact mathematical framework necessary to compute scattering analytically. 

The paper is organized as follows. In Sec.~\ref{sec:Wave_equation_derivation}, we derive the wave equation for a channel flow with constant shear and extract the modified effective metric. In Sec.~\ref{sec:Black_hole}, we analyze the stationary transcritical background, identifying the analogue horizon and deriving the modified surface gravity. In Sec.~\ref{sec:Wave_scattering}, we quantify the scattering induced by inhomogeneities in the conformal factor, demonstrating that the presence of shear actively suppresses wave scattering. Standard derivations regarding wave properties and the phase singularity at the horizon are relegated to the Appendices.

%%%%%%%%%%%%%%%%%%%%%%%%%%%%%%%%%%%%%%%%%%%%%%%%%%%%%%%%%%%%%%%%%%%%%%%%%%%%%%%%%%%%%%%%%%%%%%%%%%%%%%%%%%%%%%%%%%%%%%%%%
\section{Effective Metric for a Constant-Shear Flow}\label{sec:Wave_equation_derivation}

\begin{figure}
    \centering
    \includegraphics[width=0.5\columnwidth]{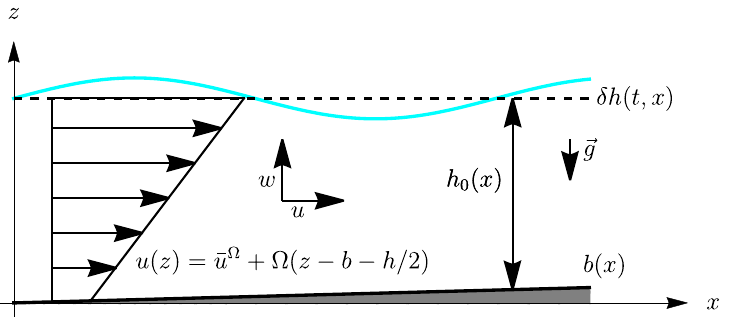}
    \caption{Schematic representation of the physical system. A two-dimensional shallow water flow propagates over a fixed channel bed $b(x)$. The unperturbed water depth is denoted by $h_0(x)$, while $\delta h(t,x)$ represents the surface wave perturbation. Crucially, the background flow exhibits a linear vertical velocity profile $u(z)$, characterized by a constant transverse shear vorticity $\Omega$.}
    \label{fig:system}
\end{figure}

In this section, we derive the wave equation for surface waves in the presence of a constant background shear, a configuration not yet explored in the Analogue Gravity context~\cite{PhysRevD.66.044019,churilov2019scattering}. In Fig.~\ref{fig:system} we report a scheme of the system under consideration. We work in the long-wavelength (shallow-water) limit, which is typically described by the Saint-Venant equations and characterized by the small parameter $kh_0 \ll 1$, where $k$ is the wave number and $h_0$ is the water depth. Furthermore, since the equations will be linearized, we assume small wave amplitudes, $ka \ll 1$, where $a$ is the wave amplitude. In this linear, non-dispersive regime, we expect to recover an equation of the form of Eq.~\eqref{eq:intro_KG}. By introducing a constant, non-zero vorticity, we present an original development of this standard framework.

\subsection{Basic equations\label{sec:Basic_equations}}

We begin by specifying the complete set of fundamental equations describing our system. We let $x$ be the longitudinal coordinate along the flume, $y$ the transverse coordinate, and $z$ the vertical coordinate (required here to describe the flow beneath the free surface, in contrast to Eq.~\eqref{eq:intro_KG} which is meant to describe waves {\it on} the free surface). The flow is assumed to have no dependence on, and no velocity component along, $y$. We thus consider a {\it two-dimensional} flow, such as might be found in an elongated water channel. 

\subsubsection{Equations in the bulk}

We assume an {\it inviscid} flow, subject to no viscous damping or friction at the boundaries. Such a flow is governed by the Euler equations~\cite{landau_fluid_2013}
\begin{equation}
    \frac{d\textbf{v}}{dt}=\partial_t\textbf{v}+(\textbf{v}\cdot\boldsymbol{\nabla})\textbf{v}=-\frac1\rho\boldsymbol{\nabla} P+\textbf{g} \, ,
    \label{eq:euler}
\end{equation}
where $\textbf{v} = u\, \hat{\textbf{x}} + w\, \hat{\textbf{z}}$ is the local velocity of the fluid, $P$ is the pressure, $\rho$ the density, and $\textbf{g}=-g\hat{\textbf{z}}$ is the acceleration due to gravity.

We assume the flow to be {\it incompressible}, with the density $\rho$ remaining constant and uniform throughout the fluid. The continuity equation for mass conservation then becomes a constraint on the flow velocity $\textbf{v}$: 
\begin{equation}
    \boldsymbol{\nabla}\cdot\textbf{v}=\partial_x u+\partial_z w=0.
    \label{eq:incompressibility}
\end{equation}

We assume that the flow possesses a constant, non-vanishing \emph{vorticity}, defined as $\bf \Omega= \nabla \times {\bf v}$.
As our flow is in the two-dimensional plane $(x,z)$, the vector ${\bf \Omega}$ always points in the transverse direction: ${\bf\Omega}=\Omega\,\hat{\bf y}$.  The vorticity can thus be treated as a scalar: $\Omega = {\bf \Omega} \cdot \hat{\bf y} = \partial_{z}u-\partial_{x}w$.
Applying the curl to the Euler equation~\eqref{eq:euler}, the governing equation for the vorticity yields:
\begin{equation}
    \left(\partial_{t}+{\bf v}\cdot\nabla\right)\Omega = 0 \,.
    \label{eq:vorticity_equation_2D}
\end{equation}
This is trivially satisfied in a flow with constant $\Omega$. We restrict our attention to this case in what follows.

An important point to raise here is that, although we will set $\Omega$ to a constant value so that it becomes a parameter of the fixed background flow, this does not automatically imply that the {\it total} vorticity will be constant. We also wish to consider perturbations on top of this background which, {\it a priori}, may carry some non-zero vorticity $\delta\Omega$. However, when linearizing Eq.~(\ref{eq:vorticity_equation_2D}) to get the governing equation for the vorticity of the perturbation, we find simply:
\begin{equation}
    \left(\partial_{t}+{\bf v}_{0}\cdot\nabla\right) \delta\Omega = 0 \,,
\end{equation}
where ${\bf v}_{0}$ is the background flow velocity. This implies that any non-zero perturbation of the vorticity will simply be advected by the flow, remaining motionless in the local rest frame of the fluid (and is thus best considered as a perturbation of the background flow itself). As we are interested in propagating wave solutions, we remove this possibility by restricting our analysis to the case $\delta\Omega = 0$ when linearizing the equations below.

\subsubsection{Boundary conditions}

The boundaries of the flow are defined as follows:
\begin{itemize}
    \item the bottom, which we shall assume fixed: $z = b(x)$, representing some arbitrary but fixed shape of the channel bed; and
    \item the free surface, which can be written as $z = b(x) + h(t,x)$; $h$ is then the instantaneous depth of the fluid at a particular position.
\end{itemize}
At each boundary, the flow velocity can be decomposed into components tangent and orthogonal to the boundary, where the orthogonal component is accompanied by a change in the boundary's position over time. Since the bottom is assumed fixed, the no-penetration condition requires that the orthogonal velocity component there must vanish. At the free surface, it will engender a time-derivative of $h$. We thus have the following {\it kinematical} boundary conditions:
\begin{eqnarray}
    \left[ w - u \, \partial_{x}b \right]_{z=b} &=& 0 \,, \nonumber \\
    \left[ w - u \, \partial_{x}\left(b+h\right) \right]_{z=b+h} &=& \partial_{t}h \,.
    \label{eq:BC_kinematic}
\end{eqnarray}
Finally, there is a {\it dynamical} boundary condition enforcing the continuity of pressure at the free surface. For simplicity, we may neglect any variations in air pressure and set it to a constant atmospheric value, which in turn may be set to zero because only pressure gradients enter the governing equations. Thus, our final boundary condition is:
\begin{equation}
    \left[ P \right]_{z=b+h} = 0 \,.
    \label{eq:BC_pressure}
\end{equation}

\subsubsection{Effective continuity equation}

Let us define the depth-averaged velocity:
\begin{equation}
    \bar{u}^\Omega(t,x) = \frac{1}{h(x)} \int_{b(x)}^{b(x)+h(t,x)} u(t,x,z) \, dz \,.
\end{equation}
We indicate the average speed of the current with the superscript $\Omega$ to highlight its dependence on the background shear and to emphasize the departure from the irrotational case.
Through a combination of the incompressibility condition~(\ref{eq:incompressibility}) and the kinematical boundary conditions~(\ref{eq:BC_kinematic}), it is straightforward to show that:
\begin{equation}
    \partial_{t}h + \partial_{x}\left(h \, \bar{u}^\Omega\right) = 0 \,.
    \label{eq:continuity_H}
\end{equation}
This is intuitive: since the flow is incompressible with a constant 3D density, $h$ plays the role of the local 1D ``density'' of the fluid, while $h\,\bar{u}^\Omega$ represents the local ``current''. Equation~(\ref{eq:continuity_H}) is simply an expression of the local conservation of the fluid's volume.

Equation~(\ref{eq:continuity_H}) is one of the two governing equations for the fields $\bar{u}^\Omega$ and $h$. It must be complemented by an equation for $\partial_{t}\bar{u}^\Omega$. To obtain such an equation in a tractable form, we first apply a slowly-varying approximation.

\subsection{Slowly-varying limit\label{sec:LWL}}

We now take the limit in which variations in the longitudinal ($x$) direction and in time occur on much longer scales than variations in the vertical ($z$) direction. To implement this, we introduce a dimensionless scaling parameter $\varepsilon \ll 1$ that acts as if the system were being stretched along $x$. We make the following replacements:
\begin{equation}
    \label{eq:lwl}
    \partial_t,\ \partial_x,\ w\ \to\ \varepsilon\partial_t,\ \varepsilon\partial_x,\ \varepsilon w \,.
\end{equation}
These are then substituted into the previous equations, and we retain only the lowest-order terms in $\varepsilon$. In this way, we focus on the long-wavelength, long-period behavior, and the short-scale physics associated with the extra dimension $z$ is integrated out. It is important to emphasize here that the first dispersive corrections occur for $kh \sim 1$, where $h$ is the depth of the fluid. Hence, we take a strict limit in which $kh \ll 1$.

This procedure simplifies two of the governing equations. First, the vorticity now takes the form:
\begin{equation}
    \Omega = \partial_{z}u \,,
    \label{eq:vorticity_LWL}
\end{equation}
with no remaining contribution from the vertical velocity component $w$.  
Since we assume $\Omega$ is strictly constant, this implies that, at a fixed $x$, $u$ varies linearly with $z$:
\begin{equation}
    u =  u^{(b)} + \Omega (z-b)=u^{(s)}+\Omega(z-h-b) \,,
    \label{eq:linear_shear_flow}
\end{equation}
where $u^{(b)}$ and $u^{(s)}$ are the current speeds at the bottom $\left.u\right|_{z=b}$ and at the free surface $\left.u\right|_{z=b+h}$, respectively.
Equation~(\ref{eq:linear_shear_flow}) implies that the velocity differential across the fluid layer depends directly on the vorticity $\Omega$:
\begin{equation}
    u^{(s)}-u^{(b)} = \Omega h \,,
\end{equation}
while the depth-averaged velocity entering Eq.~(\ref{eq:continuity_H}) becomes:
\begin{equation}
    \bar{u}^\Omega = \frac{1}{2}\left(u^{(b)}+u^{(s)}\right) = u^{(b)}+\frac{1}{2}\Omega h= u^{(s)}-\frac{1}{2}\Omega h \,.
    \label{eq:depth-averaged_velocity}
\end{equation}
Note that, in the irrotational limit ($\Omega\to0$), we recover the standard plug flow profile where the current speed does not depend on the vertical coordinate $z$:
\begin{equation}
    \lim_{\Omega\to0}\bar{u}^\Omega=\bar{u}^0=u_{\rm plug}\,,\qquad u^{(s)}-u^{(b)}=0\,.
\end{equation}

The second change effected by the slowly-varying limit occurs in the $z$-component of the Euler equations~(\ref{eq:euler}), which reduces to:
\begin{equation}
    \frac{1}{\rho} \partial_{z}P + g = 0 \,.
\end{equation}
This indicates that the pressure is strictly hydrostatic. Given the boundary condition~(\ref{eq:BC_pressure}), we immediately find:
\begin{equation*}
P = \rho g \left(b+h-z\right) \,.
\end{equation*}
Substituting this hydrostatic pressure and Eq.~(\ref{eq:vorticity_LWL}) back into the $x$-component of the Euler equations yields:
\begin{equation}
    \partial_{t}u + \partial_{x}\left[\frac{u^{2}}{2} + g\left(h+b\right)\right] + \Omega \, w = 0 \,.
    \label{eq:Euler-x}
\end{equation}
Since both $u$ and $w$ depend on $z$, we must express this equation in terms of $\bar{u}^\Omega$ to obtain an effective 1D equation analogous to Eq.~(\ref{eq:continuity_H}). By evaluating it at the fixed bottom boundary ($z=b$), where $w$ is entirely determined by $u$ due to the boundary condition~(\ref{eq:BC_kinematic}), we can combine the terms:
\begin{eqnarray}
    \left[\partial_{x}\left(\frac{u^{2}}{2}\right) + \Omega \, w\right]_{z=b} &=& \left[u \left(\partial_{x}u + \partial_{z}u \, \partial_{x}b\right)\right]_{z=b} \nonumber \\
    &=& u\left(x,z=b\right) \, \frac{d}{dx}\left[ u\left(x,z=b\right) \right] \nonumber \\
    &=& \partial_{x}\left(\frac{u^{(b) 2}}{2}\right) \,.
\end{eqnarray}
Equation~(\ref{eq:Euler-x}) then simplifies to:
\begin{equation}
    \partial_{t}u^{(b)} + \partial_{x}\left[\frac{u^{(b)2}}{2} + g\left(h+b\right)\right] = 0 \,.
\end{equation}
To express this entirely in terms of the effective flow velocity $\bar{u}^\Omega$, we substitute $\bar{u}^\Omega=u^{(b)}+\Omega h/2$ into the above equation and use Eq.~(\ref{eq:continuity_H}) to eliminate the time-derivative of $h$. This procedure yields the modified momentum equation:
\begin{equation}
    \partial_{t}\bar{u}^\Omega + \partial_{x}\left[\frac{(\bar{u}^\Omega)^{2}}{2} + g\left(h+b\right) + \frac{\Omega^{2} h^{2}}{8} \right] = 0 \,.
    \label{eq:continuity_u_constantOmega}
\end{equation}
Equations~(\ref{eq:continuity_H}) and~(\ref{eq:continuity_u_constantOmega}) represent the primary governing equations of our system. The dynamics are fully characterized by the one-dimensional fields $h(t,x)$ and $\bar{u}^\Omega(t,x)$, with $b(x)$ representing an externally imposed channel constraint and $\Omega$ serving as a fixed parameter of the background flow.

\subsection{Linearization\label{subsec:Linearization}}

The Analogue Gravity viewpoint relies on our ability to separate the system into two components of very different strengths~\cite{PhysRevD.110.116022}.  One is the strong background, which plays the role of the effective spacetime.  The other is the perturbations on top of this background, which are supposed to be sufficiently small that they do not affect the background significantly.  This is equivalent to the idea of a ``test particle'' in gravity theory, which responds to the presence of a gravitational field but whose contribution to the gravitational field need not be taken into account~\cite{wald2010general,thorne2000gravitation}.

We decompose the variables into a time-independent background flow (subscript $0$) and a time-dependent perturbation of very small amplitude (prefix $\delta$):
\begin{equation}
    h(t,x) = h_{0}(x) + \delta h(t,x) \,, \qquad \bar{u}^\Omega(t,x) = \bar{u}_{0}^\Omega(x) + \delta\bar{u}^\Omega(t,x) \,.
\end{equation}
By plugging these into Eqs.~(\ref{eq:continuity_H}) and~(\ref{eq:continuity_u_constantOmega}) and linearizing with respect to the perturbations $\delta h$ and $\delta\bar{u}^\Omega$, we isolate the equations governing the wave dynamics on the fixed background:
\begin{eqnarray}
    \left(\partial_{t}+\partial_{x}\bar{u}_{0}^\Omega\right)\delta h + \partial_{x}\left(h_0\,\delta\bar{u}^\Omega\right) &=& 0 \,, \label{eq:continuity_h_linearized} \\
    \left(\partial_{t}+\partial_{x}\bar{u}^{\Omega}_0\right)\delta\bar{u}^\Omega + \partial_{x}\left[ \left(g + \frac{\Omega^{2}h_{0}}{4}\right) \delta h \right] &=& 0 \,. \label{eq:continuity_u_linearized}
\end{eqnarray}

Because the dynamical fields $\delta h$ and $\delta \bar{u}^\Omega$ depend only on $t$ and $x$, they can be expressed via a potential. In standard irrotational flows, one introduces a velocity potential such that ${\bf u_{\rm plug}} = -\nabla\varphi$. Here, we similarly define $\delta\bar{u}^\Omega = -\partial_{x}\left(\delta\varphi^\Omega\right)$. Integrating Eq.~(\ref{eq:continuity_u_linearized}) with respect to $x$ allows us to write the system as:
\begin{eqnarray}
    \left(\partial_{t}+\partial_{x}\bar{u}_{0}^\Omega\right) \delta h - \partial_{x}\left(h_{0}\,\partial_{x}\delta\varphi^\Omega\right) &=& 0 \,, \\
    -\left(\partial_{t}+\bar{u}_{0}^\Omega\partial_{x}\right)\delta\varphi^\Omega + \left(g + \frac{\Omega^{2}h_{0}}{4}\right) \delta h &=& 0 \,.
\end{eqnarray}
These expressions can be combined into a single second-order wave equation for the potential $\delta\varphi^\Omega$:
  \begin{equation}
    \left[ \left(\partial_{t}+\partial_{x}\bar{u}_{0}^\Omega\right)\frac{1}{g+\frac{\Omega^{2}h_{0}}{4}}\left(\partial_{t}+\bar{u}_{0}^\Omega\partial_{x}\right) - \partial_{x}h_{0}\partial_{x} \right] \delta\varphi^\Omega = 0 \,.
    \label{eq:wave_eqn_on_shear_flow}
\end{equation}
This wave equation is structurally identical to Eq.~\eqref{eq:intro_KG}, which governs the propagation of a massless scalar field in the ``flowing'' effective spacetime metric~\eqref{eq:intro_metric}. This means that, even with the inclusion of a constant shear, we recover the formal analogy between a scalar field on a curved Painlevé-Gullstrand spacetime and surface waves on a flowing fluid. By comparing the two equations, we explicitly read off the modified background velocity $\bar{u}_{0}^\Omega$, the local wave speed $c_\Omega$, and the conformal factor $\Lambda_\Omega$:
\begin{equation}
    \bar{u}_{0}^\Omega = u_0^{(s)}-\Omega\frac{h_0}{2} \,, \qquad c_\Omega^{2} = g h_{0} + \frac{\Omega^{2}h_{0}^{2}}{4}=c_{\Omega=0}^2\left(1+\frac{\Omega^2h_0}{4g}\right) \,, \qquad \Lambda_\Omega^{2} = \frac{h_{0}/h_{\rm up}}{1+\frac{\Omega^{2}h_{0}}{4g}} =\frac{\Lambda^2_{\Omega=0}}{1+\frac{\Omega^2h_0}{4g}} \,.
    \label{eq:metric_parameters}
\end{equation}
Here, $c_{\Omega=0}=\sqrt{gh_0}$ and $\Lambda_{\Omega=0}=\sqrt{h_0/h_{\rm up}}$ denote the expressions defined for standard irrotational setups~\cite{PhysRevD.66.044019,Euv__2020}, and $h_{\rm up}$ is the upstream water depth used to adimensionalize the conformal factor. The expression derived for $c_\Omega$ agrees perfectly with the long-wavelength limit of the dispersion relation found in previous fluid mechanics literature~\cite{maïssa2013influenceshearflowvorticitywavecurrent,maissa2016negative,maissa2016wave}. It is important to note that the background water depth $h_0$ implicitly depends on the constant background vorticity $\Omega$, which further increases the departure from the irrotational case. We defer the detailed study of these implicit background corrections to Sec.~\ref{sec:Black_hole}, as the primary purpose of the present section is to establish the explicit rotational corrections to the governing equations and the effective metric parameters.

As detailed in Appendix~\ref{app:wave_properties}, inserting a plane wave ansatz into Eq.~\eqref{eq:wave_eqn_on_shear_flow} immediately yields the long-wavelength dispersion relation $\omega = (\bar{u}_0^\Omega \pm c_\Omega)k$. This confirms that $c_\Omega$ acts as the intrinsic phase velocity of the waves, while $\bar{u}_0^\Omega$ provides the standard convective Doppler shift. The $\pm$ sign identifies the \emph{co-current} ($+$) and \emph{counter-current} ($-$) modes, \emph{i.e.} the modes which propagate {\it with} and {\it against} the current, respectively. A comprehensive analysis of the associated physical wave properties, including momentum, wave action, and the emergence of negative-energy modes in supercritical flows, is relegated to Appendix~\ref{app:wave_properties}.

To precisely quantify the divergence from the irrotational case, we define the Brard number, ${\rm Br}$:
\begin{equation}
    {\rm Br} = \frac{\Omega (\Omega h_0)}{4g} = \frac{\Omega^2 h_0^2}{4 g h_0}\,.
\end{equation}
This dimensionless number compares the rotational correction term to the standard irrotational gravity wave speed ($c_{\Omega=0}^2 = g h_0$). For ${\rm Br} \ll 1$, the flow reduces to the irrotational limit. Conversely, for ${\rm Br} \gg 1$, rotational vorticity effects dominate the system dynamics, causing the conformal factor $\Lambda_\Omega^2$ to become nearly independent of $h_{0}$ and hence of position. This critical modification alters the wave scattering dynamics, which is governed by the conformal factor, as explored in Sec.~\ref{sec:Wave_scattering}.

Finally, using the conserved background flux $q=\bar{u}_0^\Omega h_0$ (a more formal introduction will be given in Sec.~\ref{subsec:bg}), we define a generalized Froude number for the constant-shear flow:
\begin{equation}
    {\rm Fr_\Omega} = \frac{|\bar{u}_0^\Omega|}{c_\Omega} = \frac{q}{\sqrt{g h_{0}^{3} + \frac{1}{4}\Omega^{2}h_{0}^{4}}} = \frac{{\rm Fr}_{\Omega = 0}}{\sqrt{1+{\rm Br}}} \,.
\end{equation}
This parameter retains its crucial physical role: it distinguishes subcritical (${\rm Fr_\Omega} < 1$, or $|\bar{u}_0^\Omega|<c_\Omega$) from supercritical regions (${\rm Fr_\Omega} > 1$, or $|\bar{u}_0^\Omega|>c_\Omega$), with the analogue event horizon strictly occurring at ${\rm Fr_\Omega} = 1$ (or $|\bar{u}_0^\Omega|=c_\Omega$). In Appendix~\ref{sec:Wave_energy}, we show the explicit calculation for the wave energy and we formally demonstrate that the subcritical and supercritical regimes present, as for the irrotational case, a counter-current mode which positive and negative energy, respectively.

%%%%%%%%%%%%%%%%%%%%%%%%%%%%%%%%%%%%%%%%%%%%%%%%%%%%%%%%%%%%%%%%%%%%%%%%%%%%%%%%%%%%%%%%%%%%%%%%%%%%%%%%%%%%%%%%%%%%%%%%%
\section{Transcritical Background and Surface Gravity}\label{sec:Black_hole}

Having established the modified effective metric, we now examine the stationary background with a constant shear flow, since it is not \textit{a priori} obvious how to construct an analogue black hole in the presence of vorticity. We are specifically interested in transcritical flows, which pass from a subcritical region (playing the role of the ``black hole exterior'') to a supercritical one (the analogue of the ``black hole interior''), thereby forming an analogue event horizon.

\subsection{Stationary black-hole flow with constant shear}\label{subsec:bg}
We treat the background as a {\it stationary} flow. Therefore, the equations describing it are found by setting all time derivatives in the governing equations to zero. Equations~(\ref{eq:continuity_H}) and~(\ref{eq:continuity_u_constantOmega}) then immediately yield
\begin{equation}
    \partial_{x} \left[ \bar{u}_0^\Omega \, h_{0} \right] = 0 \,, \qquad
    \partial_{x} \left[ \frac{(\bar{u}_0^\Omega)^{2}}{2g} + \frac{\Omega^{2}h_{0}^{2}}{8g} + h_{0} + b \right] = 0 \,,
\end{equation}
where $\bar{u}_0^\Omega$ is defined in Eqs.~\eqref{eq:metric_parameters}.
The stationary background flow is thus characterized by two conserved quantities. One is the {\it flow rate}, $q = \bar{u}_0^\Omega h_{0}$. We may write $\bar{u}_0^\Omega = q/h_{0}$ and plug this into the second conservation equation, which can now be expressed solely in terms of the water depth $h_{0}$:
\begin{equation}
    e_\Omega\left(h_{0}\right) + b = {\rm const.} \,, \qquad e_\Omega\left(h_{0}\right) = \frac{q^{2}}{2gh_{0}^{2}} + \frac{\Omega^{2} h_{0}^{2}}{8g} + h_{0} \,.
\end{equation}
Up to a term proportional to $q$, $e_\Omega(h_{0})+b$ may be thought of as the specific energy of a fluid parcel on the surface of the flow. Restricting to positive ({\it i.e.}, physical) values of $h_{0}$, the function $e_\Omega\left(h_{0}\right)$ has a single minimum and goes to infinity for $h_{0} \to 0$ and $h_{0} \to \infty$. The minimum occurs at the zero of the derivative:
\begin{equation}
    e_\Omega^{\prime}\left(h_{0}\right) = -\frac{q^{2}}{gh_{0}^{3}} + \frac{\Omega^{2}h_{0}}{4g} + 1 = \frac{1}{gh_{0}}\left(-\frac{q^{2}}{h_{0}^{2}} + gh_{0} + \frac{\Omega^{2}h_{0}^{2}}{4} \right) = \frac{1}{gh_{0}}\left(-(\bar{u}_0^\Omega)^{2}+c_\Omega^{2}\right) \,.
    \label{eq:eprime}
\end{equation}
Therefore, $e\left(h_{0}\right)$ achieves its minimum value wherever the effective flow velocity is equal in magnitude to the wave speed. For a given value of $q$, this defines a unique critical water depth $h_{\rm crit}$, which implicitly depends on the vorticity. From Eq.~\eqref{eq:eprime}, we can infer that for a fixed $q$, $h_{\rm crit}$ decreases as $\Omega$ increases; indeed, for strongly rotational flows (${\rm Br} \gg 1$), we have $h_{\rm crit} \sim \sqrt{q/\Omega}$. For $h_{0} > h_{\rm crit}$, $e_\Omega\left(h_{0}\right)$ is an increasing function of $h_{0}$, meaning $(\bar{u}_0^\Omega)^{2} < c_\Omega^{2}$; we refer to this as a {\it subcritical} flow. Conversely, for $h_{0} < h_{\rm crit}$, $e_\Omega\left(h_{0}\right)$ is a decreasing function of $h_{0}$, meaning $(\bar{u}_0^\Omega)^{2} > c_\Omega^{2}$, which we refer to as a {\it supercritical} flow.  

While $q$ and $b(x)$ are experimentally controlled, the value of $e_\Omega+b$ is determined by the global flow conditions. Here, we are most interested in transcritical flows. For the transition between subcritical and supercritical regimes to occur smoothly, the fluid must pass through the critical depth $h_{\rm crit}$ exactly where the specific energy requires it. Since $e_\Omega(h_{0})+b$ is constant and $e_\Omega(h_{\rm crit}) = e_{\Omega,\rm min}(q,\Omega)$ is the absolute minimum possible value of $e_\Omega(h_{0})$, this transition can only occur where the obstacle $b(x)$ achieves its {\it maximum} value, $b_{\rm max}$. Therefore, for a transcritical flow,
\begin{equation}
    e_\Omega\left(h_{0}\right) + b(x) = e_\Omega\left(h_{\rm crit}\right) + b_{\rm max} = e_{\Omega,\rm min}(q,\Omega)+b_{\rm max} \,.
\end{equation}
The depth profile is thus fully determined by the values of $q$, $\Omega$, and $b_{\rm max}$. As previously stated, for fixed $q$ and $b_{\rm max}$, the depth profile $h_{0}(x)$ inherently depends on the vorticity. The analogue horizon, where the generalized Froude number ${\rm Fr_\Omega}=1$, is strictly pinned to the top of the obstacle, where $b=b_{\rm max}$.

\subsection{Modified surface gravity}

Before evaluating the surface gravity, we must briefly recall the physical mechanism of analogue pair creation. The horizon splits the flow into subcritical ($|\bar{u}_0^\Omega| < c_\Omega$) and supercritical ($|\bar{u}_0^\Omega| > c_\Omega$) regions. In the supercritical region (the interior of the analogue black hole), the background current physically overwhelms the intrinsic wave speed. Consequently, a counter-propagating wave attempting to travel upstream is dragged inexorably downstream. As detailed in Appendix~\ref{app:wave_properties}, this kinematic reversal causes the wave's laboratory frequency $\omega$ and effective co-moving frequency $\omega^{(0)}=\omega-\bar{u}_0^\Omega k$ to carry opposite signs, forcing the wave's energy in the laboratory frame to become strictly negative. The horizon therefore permits the stimulated, energy-conserving emission of a positive-energy wave directed upstream and a negative-energy wave swept downstream.

The continuous emission of these analogue Hawking pairs is mathematically governed by the phase singularity encountered by these counter-propagating modes exactly at the horizon. The full derivation (provided in Appendix~\ref{app:hawking}) demonstrates that the relative amplitude of the emitted negative-energy waves, denoted by $\beta$, rigorously follows a thermal Planck distribution:
\begin{equation}
    \left|\beta\right|^{2} = \frac{1}{e^{2\pi \omega/\kappa_\Omega}-1} \,, \qquad \text{with} \qquad k_{B}T_H^\Omega = \frac{\hbar\kappa_\Omega}{2\pi} \,.
\end{equation}
Here, $T_H^\Omega$ is the analogue Hawking temperature and $\kappa_\Omega$ is the effective modified surface gravity, both parameters given for a constant-shear flow of vorticity $\Omega$. For our effective metric, this surface gravity is determined by the spatial gradient of the characteristic velocities evaluated exactly at the horizon:
\begin{equation}
    \kappa_\Omega = \left| \partial_{x}\left(|\bar{u}_0^\Omega|-c_\Omega\right)\right|_{\rm hor} = \left| \frac{1}{2 c_\Omega} \partial_{x}\left((\bar{u}_0^\Omega)^{2}-c_\Omega^{2}\right) \right|_{\rm hor} \,.
    \label{eq:kappa_definition}
\end{equation}
This prediction applies to any system capable of realizing an analogue black-hole horizon. Let us apply it to the transcritical shear flow described above to determine exactly how background vorticity affects the emission temperature.

Both $\bar{u}_0^\Omega$ and $c_\Omega$ depend on the water depth $h_{0}$, so we need to know how the derivative $dh_{0}/dx$ behaves at the horizon. Differentiating the constancy equation $e_\Omega\left(h_{0}\right) + b = \text{const.}$ with respect to $x$ yields:
\begin{equation}
    e_\Omega^{\prime}\left(h_{0}\right) \, \frac{dh_{0}}{dx} + \frac{db}{dx} = 0 \,.
    \label{eq:dhdx}
\end{equation}
At the horizon, we have established that $e_\Omega^{\prime}\left(h_{\rm crit}\right) = 0$ and $db/dx = 0$, so Eq.~(\ref{eq:dhdx}) is satisfied even though $dh_{0}/dx$ remains finite. Differentiating a second time, we have
\begin{equation}
    e_\Omega^{\prime\prime}\left(h_{0}\right) \, \left( \frac{dh_{0}}{dx} \right)^{2} + e_\Omega^{\prime}\left(h_{0}\right) \, \frac{d^{2}h_{0}}{dx^{2}} + \frac{d^{2}b}{dx^{2}} = 0 \,,
\end{equation}
and evaluated at the horizon, this reduces to
\begin{equation}
    \left|\frac{dh_{0}}{dx}\right|_{\rm hor} = \sqrt{\frac{\left.-d^{2}b/dx^{2}\right|_{\rm hor}}{e_\Omega^{\prime\prime}\left(h_{\rm crit}\right)}} \,, \qquad e_\Omega^{\prime\prime}\left(h_{\rm crit}\right) = \frac{3q^{2}}{gh_{\rm crit}^{4}} + \frac{\Omega^{2}}{4g} \,.
\end{equation}
Using Eq.~(\ref{eq:eprime}) to substitute $(\bar{u}_0^\Omega)^{2}-c_\Omega^{2} = -gh_{0} \, e_\Omega^{\prime}\left(h_{0}\right)$ into our definition of surface gravity (Eq.~\ref{eq:kappa_definition}), we obtain:
\begin{equation}
    \kappa_\Omega = \left| \frac{h_{0}}{2q} \, gh_{0} \, e_\Omega^{\prime\prime}\left(h_{0}\right) \, \frac{dh_{0}}{dx}  \right|_{\rm hor} = \frac{gh_{\rm crit}^{2}}{2q} \sqrt{e_\Omega^{\prime\prime}\left(h_{\rm crit}\right)} \sqrt{\left.-\frac{d^{2}b}{dx^{2}}\right|_{\rm hor}} \,.
\end{equation}
Plugging in $e_\Omega^{\prime\prime}\left(h_{\rm crit}\right)$ and imposing $e_\Omega^{\prime}\left(h_{\rm crit}\right) = 0$ in Eq.~(\ref{eq:eprime}), this simplifies to:
\begin{equation}
    \kappa_\Omega = \sqrt{1-\frac{gh_{\rm crit}^{3}}{4q^{2}}} \, \sqrt{-g\left.\frac{d^{2}b}{dx^{2}}\right|_{\rm hor}}=\sqrt{1-\frac{1}{4({\rm Br_{crit}}+1)}}\sqrt{-g\left.\frac{d^{2}b}{dx^{2}}\right|_{\rm hor}} \,,
    \label{eq:surface_gravity}
\end{equation}
where we have used the definition of Brard number evaluated at the horizon, $\rm Br|_{hor}=Br_{crit}=\frac{\Omega^2h_{\rm crit}}{4g}$.
The prefactor $\sqrt{1-gh_{\rm crit}^{3}/4q^{2}}$ dictates how $\kappa_\Omega$ depends on the parameters of the flow that are independent of the obstacle geometry, specifically $\Omega$ and $q$ (noting that $h_{\rm crit}$ is a function of both via Eq.~(\ref{eq:eprime})). We also note that when the prefactor is expressed in terms of the critical Brard number, $\rm Br_{crit}$, the explicit dependence on the flow rate $q$ vanishes.

When the flow is irrotational ($\Omega=0$), we have $gh_{\rm crit}^{3}/q^{2}=1$ (or $\rm Br_{crit}=0$), and $\kappa_{\Omega=0}$ depends purely on the geometrical curvature of the obstacle at the horizon, $\sqrt{-g\, d^2 b/dx^2|_{\rm hor}}$~\cite{bossa2025}. Thus, for an irrotational flow, the surface gravity is entirely independent of the flow rate.

\begin{figure}
    \centering
    \includegraphics[width=0.4\columnwidth]{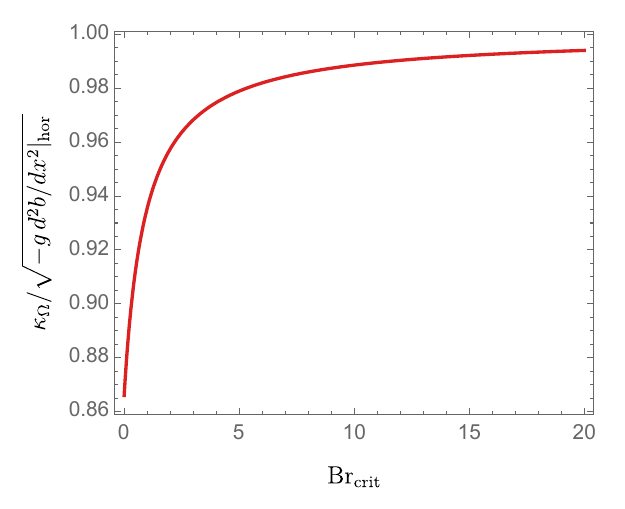}
    \caption{The prefactor of the modified surface gravity, $\kappa_\Omega / \sqrt{-g\left.\,d^2b/dx^2\right|_{\rm hor}}$, plotted as a function of the dimensionless Brard number evaluated at the horizon, ${\rm Br|_{hor}= Br_{crit}}$. As demonstrated analytically, expressing the prefactor in terms of the critical Brard number absorbs any explicit dependence on the background flow rate $q$, yielding a unique profile: $\sqrt{1 - \frac{1}{4({\rm Br_{crit}}+1)}}$. This prefactor increases monotonically from its irrotational limit of $\sqrt{3}/2 \approx 0.866$ at ${\rm Br_{crit}} = 0$ to an asymptotic limit of $1$ in the strongly rotational regime (${\rm Br_{crit}} \gg 1$).}
    \label{fig:hawkingradiation}
\end{figure}

Conversely, for a rotational flow ($\Omega\ne0$), Eq.~(\ref{eq:eprime}) ensures that $gh_{\rm crit}^3/q^2$ decreases as vorticity increases. This implies that the square root prefactor in Eq.~\eqref{eq:surface_gravity} strictly resides within the bounds: 
\begin{equation}
    \frac{\sqrt{3}}{2} \le \frac{\kappa_\Omega}{\sqrt{-g\left.\frac{d^2b}{dx^2}\right|_{\rm hor}}} \le 1\,,
\end{equation}
where the lower bound corresponds exactly to the irrotational case ($\kappa_{\Omega=0}$). In Fig.~\ref{fig:hawkingradiation}, we plot the prefactor of the modified surface gravity as a function of the critical Brard number, $\rm Br_{crit}$.

We conclude that the introduction of a constant-shear flow has two profound physical consequences for the analogue black hole. First, the modified surface gravity $\kappa_\Omega$ is no longer purely geometric; it couples directly to the intrinsic kinematic parameters of the fluid ($q$ and $\Omega$). Second, because the prefactor increases toward unity, the presence of background vorticity actively increases the modified surface gravity, thereby inducing an enhancement of the analogue Hawking temperature ($T_H^\Omega \propto \kappa_\Omega$). 

This result carries an astrophysical interpretation. In General Relativity, the surface gravity of a Schwarzschild black hole, $\kappa_{\rm SBH}$, is inversely proportional to its mass, $M_{\rm BH}$. Specifically, $\kappa_{\rm SBH} = c_{\rm light}^4 / (4GM_{\rm BH})$, where $c_{\rm light}$ is the speed of light in vacuum and $G$ is the gravitational constant. Therefore, increasing the surface gravity is formally equivalent to decreasing the mass of the black hole. If we define $M_\Omega$ as the effective mass of the analogue black hole formed by the sheared flow, we find:
\begin{equation}
    \kappa_\Omega \ge \kappa_{\Omega=0} \quad \Longleftrightarrow \quad M_\Omega \le M_{\Omega=0} \,.
\end{equation}
In short, while the macroscopic flow parameters have only a mild numerical effect on the prediction (up to a $\approx 15\%$ enhancement), introducing a background shear effectively ``lightens'' the analogue black hole, proving that fluid vorticity leaves a distinct and measurable thermodynamic imprint on the analogue horizon.

%%%%%%%%%%%%%%%%%%%%%%%%%%%%%%%%%%%%%%%%%%%%%%%%%%%%%%%%%%%%%%%%%%%%%%%%%%%%%%%%%%%%%%%%%%%%%%%%%%%%%%%%%%%%%%%%%%%%%%%%%
\section{Wave Scattering and Mode Mixing}\label{sec:Wave_scattering}

The analogue Hawking effect is essentially a localized scattering process induced by the phase singularity at the horizon. However, a more general, continuous scattering occurs throughout the inhomogeneous region of the flow due to spatial variations in the conformal factor $\Lambda(x)$. 

Given the generic effective metric in Eq.~\eqref{eq:intro_metric}, this continuous scattering is entirely characterized by the conformal factor $\Lambda(x)$. This term acts as the transverse component of the metric (\emph{i.e.}, the coefficient of $dy^2$). Although we restrict our analysis to waves that do not depend on $y$, local longitudinal inhomogeneities in this factor couple co-propagating and counter-propagating waves. In this way, the spatial variations of the conformal factor induce an effective potential barrier, playing a physical role identical to the Regge-Wheeler potential in General Relativity.

In Appendix~\ref{subsec:mode_mixing} we derive the generic mode-mixing amplitude equations for a wave-scattering scenario with an arbitrary conformal factor $\Lambda(x)$.
We now apply this general framework specifically to our constant-shear model. We substitute the generalized kinematic parameters previously established: $U=\bar{u}_0^\Omega$, $c=c_\Omega$, and $\Lambda=\Lambda_\Omega$. We introduce the corresponding unscattered modes:
\begin{equation}
    W_{\pm}^\Omega(x) = \frac{1}{\sqrt{\Lambda_\Omega(x)}} \, {\rm exp}\left(i\int^{x}k_{\pm}(x^{\prime}) dx^{\prime}\right) = \frac{1}{\sqrt{\Lambda_\Omega(x)}} \, {\rm exp}\left(i\omega\int^{x} \frac{dx^{\prime}}{\bar{u}_0^\Omega(x^{\prime}) \pm c_\Omega(x^{\prime})}\right) \,.
\end{equation}
The specific solution to the wave equation is then constructed as a superposition of these modes with position-dependent coefficients:
\begin{equation}
    \varphi^\Omega(x) = \mathcal{A}_{+}^\Omega(x) \, W_{+}^\Omega(x) + \mathcal{A}_{-}^\Omega(x) \, W_{-}^\Omega(x) \,.
\end{equation}
The amplitudes $\mathcal{A}_{+}^\Omega$ and $\mathcal{A}_{-}^\Omega$ are precisely the normalized amplitudes required to conserve the norm flux of the modes (see Appendix~\ref{app:wave_properties}). Substituting $U=\bar{u}_0^\Omega$, $c=c_\Omega$, and $\Lambda=\Lambda_\Omega$ into Eq.~\eqref{eq:amplitudes-general}, the differential equations governing the amplitude evolution become:
\begin{equation}
    \partial_{x}\left[ \begin{array}{c} \mathcal{A}_{+}^\Omega \\ \mathcal{A}_{-}^\Omega \end{array} \right] = \frac{\Lambda_\Omega^{\prime}(x)}{2\Lambda_\Omega(x)} \left[ \begin{array}{cc} 0 & e^{2i\omega\theta_\Omega(x)} \\ e^{-2i\omega\theta_\Omega(x)} & 0 \end{array} \right] \left[ \begin{array}{c} \mathcal{A}_{+}^\Omega \\ \mathcal{A}_{-}^\Omega \end{array} \right] \,, \qquad \theta_\Omega(x) = \int^{x} \frac{c_\Omega(x^{\prime})}{(\bar{u}_0^\Omega)^{2}(x^{\prime})-c_\Omega^{2}(x^{\prime})} \, dx^{\prime} \,.
    \label{eq:mode_amplitude_equation}
\end{equation}
This formulation provides a direct mechanism to compute the scattering coefficients. It is straightforward to show that $\left|\mathcal{A}_{+}^\Omega\right|^{2}-\left|\mathcal{A}_{-}^\Omega\right|^{2}$ is a conserved quantity, a necessary consequence of norm conservation for modes with opposite norm flux. Notice that by taking the irrotational limit ($\Omega\to0$), we perfectly recover the analytical solutions previously derived in the literature~\cite{Euv__2020}. 

An important complication arises because the phase $\theta_\Omega$ is singular wherever $(\bar{u}_0^\Omega)^{2}-c_\Omega^{2} = 0$. We have already demonstrated how this phase singularity engenders the Hawking process at the horizon. In the context of mode mixing, this singularity prevents us from integrating continuously across the horizon; we can only integrate up to points very close to it, $x = x_{\rm hor} \pm \epsilon$. For the co-propagating mode, which does not experience a phase singularity, we impose continuity by equating its value at $x=x_{\rm hor}\pm\epsilon$. For the counter-propagating modes, these points represent their ``in'' states, as they originate at the horizon and propagate outward.

For concreteness, we focus on a scenario where an incident co-propagating wave approaches the horizon from the subcritical side. This wave is largely transmitted across the horizon, but it is partially scattered into counter-propagating waves on either side (as illustrated in Figure~\ref{fig:scattering_processes} (Right)). To integrate Eq.~\eqref{eq:mode_amplitude_equation}, we impose the initial condition that the counter-propagating waves are purely products of the scattering process and have no support exactly on the horizon. Therefore, we set their amplitudes to zero at $x=x_{\rm hor}\pm\epsilon$. We then integrate outward from the horizon on both sides, and finally normalize the resulting amplitudes against the amplitude of the incident wave.

\begin{figure}
    \centering
    \includegraphics[width=0.9\columnwidth]{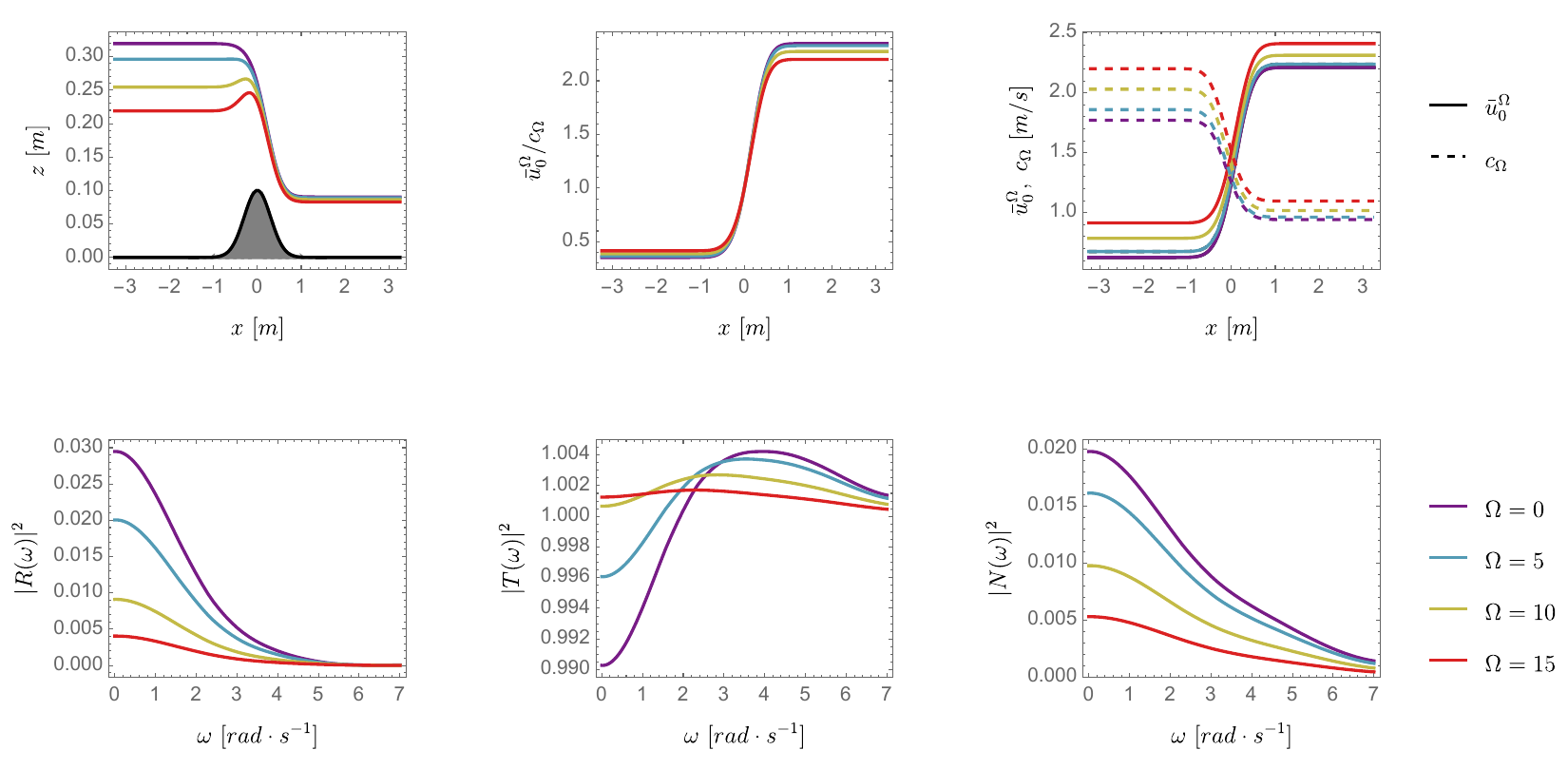}
    \caption{Scattering coefficients over a fixed Gaussian obstacle with a flow rate $q = 0.2 \, \mathrm{m}^2/\mathrm{s}$ across several values of transverse background vorticity (in units of $\mathrm{s}^{-1}$); this includes the irrotational flow with $\Omega=0$. The top row shows the background flow: the water depth (left), the Froude number (center), and the profiles of $\bar{u}_{0}^\Omega$ and $c_\Omega$ (right). On the bottom are shown the scattering coefficients for the reflected mode (left), the transmitted mode (center), and the negative-energy mode (right). We notice that the scattering is systematically suppressed for larger values of $\Omega$.}
    \label{fig:scattering_amplitudes_fixed-q}
\end{figure}

Following this procedure, we performed a numerical integration of Eq.~\eqref{eq:mode_amplitude_equation} over a fixed obstacle with a Gaussian profile (a treatment for asymmetrical obstacles is provided in Appendix~\ref{app:asymmetrical}):
\begin{equation}
    b(x) = b_{\rm max} \, {\rm exp}\left( -\frac{x^{2}}{2\sigma^{2}} \right) \,,
    \label{eq:gaussian_obstacle}
\end{equation}
with $b_{\rm max} = 10\,{\rm cm}$ and $\sigma = 30\,{\rm cm}$. We considered four different values of background vorticity: $\Omega = 0$ (irrotational flow), $5$, $10$, and $15\,\mathrm{s}^{-1}$. The flow rate was fixed at $q = 0.2\,\mathrm{m}^{2}/\mathrm{s}$. The results are plotted in Fig.~\ref{fig:scattering_amplitudes_fixed-q}. The plotted coefficients represent the squared amplitudes of the waves in the asymptotic regions:
\begin{align}
    R &= \mathcal{A}_{-}^\Omega\left(x \to -\infty\right)   &\text{(reflected wave)}\,, \nonumber \\
    T &= \mathcal{A}_{+}^\Omega\left(x \to +\infty \right)   &\text{(transmitted wave)}\,, \nonumber \\
    N &= \mathcal{A}_{-}^\Omega\left(x \to +\infty \right)   &\text{(negative-energy wave)} \,.
\end{align}
We verified that these scattering coefficients strictly satisfy the unitarity relation:
\begin{equation}
    \left|T\right|^{2}+\left|R\right|^{2}-\left|N\right|^{2} = 1 \,.
\end{equation}
Notably, when $\left|N\right|^{2}$ is larger than $\left|R\right|^{2}$, the transmission coefficient $\left|T\right|^{2}$ must exceed 1 to compensate for the production of negative-energy waves. This phenomenon is clearly visible for certain frequencies in Fig.~\ref{fig:scattering_amplitudes_fixed-q} and represents a form of energy extraction from the background flow. This is deeply analogous to the Penrose process in astrophysical black holes~\cite{Penrose-Floyd-1971}, where energy is extracted via the transfer of matter into a negative-energy state. This behaviour can be found also for irrotational flows.

\begin{figure}
    \centering
    \includegraphics[width=0.45\columnwidth]{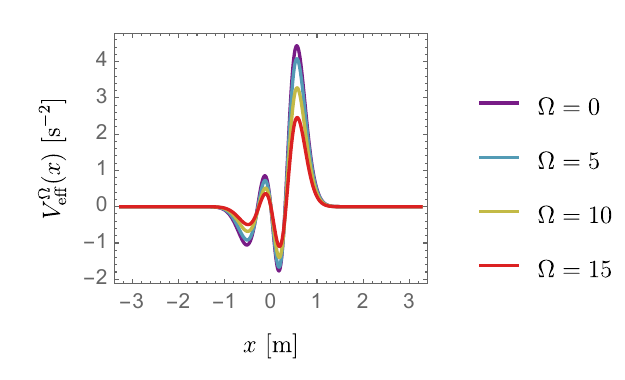}
    \caption{Spatial profile of the effective scattering potential $V_{\rm eff}^\Omega(x)$ generated by a Gaussian bottom obstacle ($b_{\rm max}=10\,{\rm cm}$, $\sigma=30\,{\rm cm}$) for a fixed flow rate $q = 0.2 \, \mathrm{m}^2/\mathrm{s}$. The curves correspond to the irrotational case ($\Omega=0$) and three constant-shear cases ($\Omega = 5, 10, 15\,\mathrm{s}^{-1}$). As the background transverse vorticity $\Omega$ increases, the peak of the effective potential barrier is significantly suppressed. This lowering of the analogue Regge-Wheeler barrier is the fundamental physical mechanism responsible for driving the greybody transmission coefficient $T$ toward unity and suppressing the production of negative-energy modes $N$, as observed in Fig.~\ref{fig:scattering_amplitudes_fixed-q}.} \label{fig:effectivepotential}
\end{figure}

The primary novelty here concerns the effect of the transverse vorticity. As $\Omega$ is increased, scattering is suppressed: both the reflection ($R$) and negative-energy ($N$) scattering coefficients decrease significantly, while the transmission coefficient ($T$) converges toward unity.
As already mentioned, the scattering between the co- and counter-propagating modes can be described by the presence of an effective potential, which acts like a Regge-Wheeler potential in an appropriate coordinate system~\cite{Anderson2014}.  An explicit derivation is given in Appendix~\ref{subsec:effectivepotential} for a massless scalar field $\phi$ propagating on the generic effective spacetime of Eq.~\eqref{eq:intro_metric}.
By substituting the specific kinematic parameters of our constant-shear flow, which are $U=\bar{u}_0^\Omega$, $c=c_\Omega$, and $\Lambda=\Lambda_\Omega$ (defined in Eq.~\eqref{eq:metric_parameters}), into the general result of Eq.~\eqref{eq:effectivepotential}, we obtain the following form for the 
effective potential in the standard coordinate system appropriate to the laboratory:
\begin{equation}
    \label{eq:effectivepotential-shear}
    V_{\rm eff}^\Omega=c_\Omega^2\left(1-\frac{(\bar{u}_0^\Omega)^2}{c_\Omega^2}\right)^2\left(\frac{\Lambda_\Omega''}{2\Lambda_\Omega}-\frac{(\Lambda_\Omega')^2}{4\Lambda_\Omega^2}\right)+\left(1-\frac{(\bar{u}_0^\Omega)^2}{c_\Omega^2}\right)\left[\left(1+\frac{(\bar{u}_0^\Omega)^2}{c_\Omega^2}\right)c_\Omega\,c_\Omega'-2\bar{u}_0^\Omega\,(\bar{u}_0^{\Omega})'\right]\frac{\Lambda_\Omega'}{2\Lambda_\Omega}\,.
\end{equation}
In Fig.~\ref{fig:effectivepotential}, we plot this effective potential $V_{\rm eff}^\Omega(x)$ for the same scattering scenarios analyzed in Fig.~\ref{fig:scattering_amplitudes_fixed-q}. As anticipated, the amplitude of the effective potential barrier strictly decreases as the background vorticity $\Omega$ increases. 

From a gravitational perspective, this behaves in the opposite way with respect to the Hawking temperature: increasing the vorticity gives an effective potential that is formally equivalent to considering an analogue black hole with a \emph{larger} effective mass.

%%%%%%%%%%%%%%%%%%%%%%%%%%%%%%%%%%%%%%%%%%%%%%%%%%%%%%%%%%%%%%%%%%%%%%%%%%%%%%%%%%%%%%%%%%%%%%%%%%%%%%%%%%%%%%%%%%%%%%%%%
%%%%%%%%%%%%%%%%%%%%%%%%%%%%%%%%%%%%%%%%%%%%%%%%%%%%%%%%%%%%%%%%%%%%%%%%%%%%%%%%%%%%%%%%%%%%%%%%%%%%%%%%%%%%%%%%%%%%%%%%%
\section{Conclusion and discussion\label{sec:Conclusion}}

In this paper, we have generalized the standard irrotational framework of Analogue Gravity to incorporate shallow water waves propagating on a constant-shear background flow. We demonstrated that a flow possessing transverse background vorticity remains perfectly compatible with the existence of an effective spacetime metric, which is not \emph{a priori} obvious. By extracting the modified kinematical parameters and the conformal factor, we showed that the underlying geometry retains a generalized Painlevé-Gullstrand structure. However, the introduction of shear fundamentally alters the analogue spacetime, yielding two opposing physical consequences for wave scattering and the analogue Hawking effect.

First, we showed that the shear flow  
modifies the 
surface gravity of the analogue black hole. By coupling the intrinsic kinematic properties of the fluid to the geometric curvature of the obstacle,
the presence of a background shear tends to increase the effective surface gravity. This corresponds to an enhancement of the predicted analogue Hawking temperature. Translated into the language of General Relativity, the introduction of fluid vorticity effectively ``lightens'' the analogue black hole, reducing its effective mass compared to the purely irrotational case.

Second, we found that the shear flow  
significantly alters the continuous wave scattering between co- and counter-propagating modes induced by the varying background. This mode-mixing process is entirely governed by the conformal factor $\Lambda_\Omega(x)$, whose spatial variations generate an analogue Regge-Wheeler potential barrier~\cite{PhysRev.108.1063} that regulates the greybody transmission coefficients. To explicitly validate this mechanism, we analytically derived the exact mathematical form of this effective scattering potential and numerically evaluated its spatial profile for our specific transcritical scattering scenarios. These numerical results directly visualize how an increasing transverse background vorticity (${\rm Br} \gg 1$) flattens the conformal factor and  
suppresses the peak of the potential barrier. 
This lowering of the analogue potential barrier naturally drives the analogue greybody transmission coefficient toward unity. 

We thus uncover a dichotomy in rotational analogue gravity: while background shear 
enhances the localized pair creation at the horizon (the Hawking effect), it simultaneously suppresses the extended spatial scattering in the surrounding curved spacetime (the greybody factors).  Both are advantageous for optimisation of the Hawking effect, for the suppression of the scattering between co- and counter-propagating modes brings the remaining scattering coefficients for the counter-propagating modes only closer to the standard Hawking prediction~\cite{Macher2009,Busch2014,Robertson2016}.

The compatibility of a constant-shear flow with the effective metric formalism provides a crucial step toward understanding real-world analogue gravity experiments, where boundary layers and shear are unavoidable. While we deliberately restricted our analysis to an idealized model with constant shear, thus maintaining  
the analytical tractability required to compute scattering over a non-flat bottom, this assumption of strictly constant vorticity throughout the entire bulk is an idealization. In future work, we will extend this velocity profile by considering a two-layer system, which will allow us to confine the vorticity to the lower layer and more accurately model realistic experimental boundary conditions. Additionally, since the phase singularity at the analogue horizon must ultimately be regularized by short-wavelength physics, the inclusion of dispersive effects (in the manner of~\cite{Unruh2013,Coutant-Parentani-2014}) on top of this sheared background remains a natural and necessary extension of this work.

%%%%%%%%%%%%%%%%%%%%%%%%%%%%%%%%%%%%%
\section*{Data availability}

The Mathematica codes associated to this work are freely available in~\cite{robertson_2026_18955484}.

%%%%%%%%%%%%%%%%%%%%%%%%%%%%%%%%%%%%%%%%%%%%%%%%%%%%%%%%%%%%%%%%%%%%%%%%%%%%%%%%%%%%%%%%%%%%%%%%%%%%%%%%%%%%%%%%%%%%%%%%%

\begin{acknowledgments}
AB and SR are funded by the CNRS Chair in Physical Hydrodynamics (reference ANR-22-CPJ2-0039-01).
This work pertains (namely, is not funded but enters in the scientific perimeter) to the French government programs ``Investissement d'Avenir'' EUR INTREE (reference ANR-18-EURE-0010) and LABEX INTERACTIFS (reference ANR-11-LABX-0017-01).
\end{acknowledgments}

%%%%%%%%%%%%%%%%%%%%%%%%%%%%%%%%%%%%%%%%%%%%%%%%%%%%%%%%%%%%%%%%%%%%%%%%%%%%%%%%%%%%%%%%%%%%%%%%%%%%%%%%%%%%%%%%%%%%%%%%%

\begin{appendices}

\renewcommand{\thesubsection}{\thesection.\arabic{subsection}}

\section{Physical Properties of Waves on a Constant-Shear Flow}\label{app:wave_properties}

In this appendix, we consider the physical properties of the wave solutions to Eq.~(\ref{eq:wave_eqn_on_shear_flow}) propagating on a constant-shear background flow. We focus primarily on the properties of plane waves; therefore, we pay particular attention to the case where the background is uniform, meaning these properties do not vary continuously in space. After deriving the dispersion relation, we calculate the momentum, energy, and norm (wave action) associated with plane waves (in the spirit of~\cite{Stepanyants-Fabrikant-1989}), noting specifically how the wave energy can become negative when the flow is supercritical. 

We demonstrate that the background vorticity $\Omega$ only enters into the definitions of the metric parameters $\bar{u}_0^\Omega$, $c_\Omega$, and $\Lambda_\Omega$. The general mathematical form of the physical quantities remains invariant under the introduction of vorticity, highlighting the robustness of the effective metric description for shear flows. (Many of these results are reproduced in Appendix~\ref{app:field_theory} from a more abstract field theory perspective).

\subsection{Dispersion relation with vorticity-dependent parameters}

Assume that the parameters $\bar{u}_0^\Omega$, $c_\Omega$, and $\Lambda_\Omega$, which have been defined in Eq.~\eqref{eq:metric_parameters} and that we recall below for simplicity:
\begin{equation}
    \bar{u}_{0}^\Omega = u_0^{(s)}-\frac{\Omega h_0}{2} \,, \qquad c_\Omega^{2} = g h_{0} + \frac{\Omega^{2}h_{0}^{2}}{4} \,, \qquad \Lambda_\Omega^{2} = \frac{h_{0}/h_{\rm up}}{1+\frac{\Omega^{2}h_{0}}{4g}} \,,
    \label{eq:metric_parameters2}
\end{equation}
are all uniform in space, and that we have a plane wave solution $\delta\varphi^\Omega \propto {\rm exp}\left(i k x - i \omega t\right)$. Inserting this into Eq.~(\ref{eq:wave_eqn_on_shear_flow}) immediately implies:
\begin{equation}
    \omega^{(0)\,2}=\left(\omega - \bar{u}_0^\Omega k\right)^{2} = c_\Omega^{2} k^{2} \qquad \iff \qquad \omega = \left(\bar{u}_0^\Omega \pm c_\Omega\right)k \,.
    \label{eq:disp_rel}
\end{equation}
This is the dispersion relation for modes on a flow with constant shear, and its real roots identify all independent plane wave solutions to the wave equation. We find two independent solutions, corresponding to one wave traveling in each direction with respect to the fluid. 

The appearance of the $-\bar{u}_0^\Omega k$ term on the left-hand side is indicative of the convective Doppler effect: the measured laboratory frequency $\omega$ differs from the {\it effective co-moving frequency} $\omega^{(0)}=\omega-\bar{u}_0^\Omega k$ that would be measured in the effective rest frame (ERF) where $\bar{u}_0^\Omega=0$. It is crucial to highlight the physical meaning behind the potential sign change of $\omega^{(0)}$. In a subcritical flow ($|\bar{u}_0^\Omega| < c_\Omega$), the counter-propagating wave successfully travels upstream, and $\omega^{(0)}$ shares the same sign as $\omega$. However, in a supercritical flow ($|\bar{u}_0^\Omega| > c_\Omega$), the background current physically overcomes the intrinsic wave speed. The wave is dragged downstream despite propagating upstream in the local fluid frame, forcing $\omega^{(0)}$ to change sign. As we will demonstrate in the following energy analysis, this kinematic sign reversal is the fundamental physical mechanism that allows the wave energy to become negative.

The total speeds of the two waves in the laboratory frame are $\bar{u}_0^\Omega \pm c_\Omega$, corresponding to the {\it co-propagating} and {\it counter-propagating} waves. This combination explicitly shows the competition between the background flow and the wave propagation. Note that only $\bar{u}_0^\Omega$ and $c_\Omega$ enter the dispersion relation, not $\Lambda_\Omega$. This is mathematically expected; as stated in the Introduction, wave kinematics are strictly determined by the classical field trajectories $ds^{2}=0$, which are unaffected by the overall conformal factor $\Lambda_\Omega$.

\subsection{Wave momentum in the presence of constant shear}

The $x$-component of the momentum density is given by $\rho \, u$, where $\rho$ is the fluid density and $u$ is the total fluid velocity in the $x$-direction. This must be integrated over space to obtain the total momentum:
\begin{equation}
    P = \int_{x_{1}}^{x_{2}} dx \int_{y_{1}}^{y_{2}} dy \int_{b}^{b+h} dz \, \rho \, u  = \rho \, W \int_{x_{1}}^{x_{2}} dx \, \bar{u}^\Omega \, h \,,
\end{equation}
where we have used the definition of the depth-averaged velocity to remove the integration over $z$, and the independence of $y$ to replace the $y$-integration by multiplication by the channel width $W$. The fields $\bar{u}^\Omega$ and $h$ will oscillate in $x$ due to the presence of waves, but these oscillations average to zero when integrated directly over a wavelength. This means that terms linear in the perturbations $\delta \bar{u}^\Omega$ and $\delta h$ can be neglected if the $x$-integral is taken as an average. Consequently, the effective momentum density of the wave is quadratic in the perturbation fields, and thus quadratic in the wave amplitude. Splitting $P$ into a background contribution and a wave contribution:
\begin{equation}
    P = \rho \, W \left\{ \int_{x_{1}}^{x_{2}} dx \, \bar{u}_0^\Omega \, h_{0} + \int_{x_{1}}^{x_{2}} dx \, \delta\bar{u}^\Omega \, \delta h \right\} \equiv P_{0} + \delta P \,.
\end{equation}

An important consideration is how the momentum transforms under a Galilean boost into another reference frame. A boost affects the velocity uniformly by the addition of a constant speed: $\bar{u}^\Omega \to \bar{u}^\Omega + V$. Because the boost velocity $V$ is constant, it is absorbed entirely into the background speed rather than the perturbation speed ({\it i.e.}, $\bar{u}_0^\Omega \to \bar{u}_0^\Omega + V$, $\delta \bar{u}^\Omega \to \delta \bar{u}^\Omega$). Thus, the perturbation velocity $\delta\bar{u}^\Omega$ is invariant under boosts, and so is the wave momentum $\delta P$.

To be explicit, let us write the potential perturbation as a plane wave of constant amplitude $A$: $\delta\varphi^\Omega = A \, {\rm sin}\left(kx-\omega t\right)$. We derive $\delta\bar{u}^\Omega$ and $\delta h$ using our previously established linearized equations:
\begin{eqnarray}
    \delta\bar{u}^\Omega = -\partial_{x}\left(\delta\varphi^\Omega\right) &=& -k A {\rm cos}\left(kx-\omega t\right) \,, \nonumber \\
    \delta h = \frac{h_{0}}{c_\Omega^{2}} \left(\partial_{t}+\bar{u}_0^\Omega\partial_{x}\right)\delta\varphi^\Omega &=& -\frac{h_{0}}{c_\Omega^{2}} \left(\omega-\bar{u}_0^\Omega k\right) A {\rm cos}\left(kx-\omega t\right) \,.
    \label{eq:plane_wave}
\end{eqnarray}
The average wave momentum per unit length is then:
\begin{equation}
    \delta P = \frac{\rho \, W}{2} \, \frac{h_{0}}{c_\Omega^{2}} \, k \, \left(\omega-\bar{u}_0^\Omega k\right) \, A^{2} \,.
\end{equation}
Notice that $\omega-\bar{u}_0^\Omega k$ is the effective frequency in the ERF (Eq.~\ref{eq:disp_rel}). It is therefore invariant under boosts, ensuring $\delta P$ is invariant as well.

\subsection{Wave energy in the presence of constant shear\label{sec:Wave_energy}}

We perform a similar analysis for the energy of the system. The local energy density is given by $\rho \left(u^{2}/2 + g z\right)$ (where, in line with the long-wavelength limit described in Sec.~\ref{sec:LWL}, we neglect the kinetic energy due to vertical motion). The total energy is:
\begin{equation}
    E = \int_{x_{1}}^{x_{2}} dx \int_{y_{1}}^{y_{2}} dy \int_{b}^{b+h} dz \, \rho \, \left( \frac{u^{2}}{2} + g z \right) = \rho \, W \, \int_{x_{1}}^{x_{2}} dx \left( \frac{(\bar{u}^\Omega)^{2}h}{2} + \frac{\Omega^{2} h^{3}}{24} + \frac{gh^{2}}{2} \right) \,,
\end{equation}
where we have used the constancy of the vorticity to write $u=\bar{u}^\Omega+\Omega\left(z-b-h/2\right)$ before integrating over $z$. As before, we split this into background and wave contributions, neglecting terms linear in the perturbations as they average to zero. Terms generating third-order perturbations also oscillate and average away. The wave energy remains quadratic in the perturbation fields:
\begin{eqnarray}
    E &=& \rho \, W \left\{ \int_{x_{1}}^{x_{2}} dx \left( \frac{(\bar{u}_0^\Omega)^{2}h_{0}}{2} + \frac{\Omega^{2} h_{0}^{3}}{24} + \frac{gh_{0}^{2}}{2} \right) + \int_{x_{1}}^{x_{2}} dx \left[ \frac{h_{0}}{2} \left(\delta\bar{u}^\Omega\right)^{2} + \bar{u}_0^\Omega \, \delta\bar{u}^\Omega \, \delta h + \left(\frac{\Omega^{2}h_{0}}{8} + \frac{g}{2}\right) \left(\delta h\right)^{2} \right] \right\} \nonumber \\
    &\equiv& E_{0} + \delta E \,.
\end{eqnarray}

Consider the transformation of the energy under a boost. As stated above, the boost velocity $V$ is absorbed into the background flow velocity $\bar{u}_0^\Omega$. However, the wave energy $\delta E$ includes a cross-term proportional to $\bar{u}_0^\Omega$. In fact, this cross-term is exactly $\bar{u}_0^\Omega$ multiplied by the wave momentum $\delta P$. Therefore, under a boost $\bar{u}_0^\Omega \to \bar{u}_0^\Omega + V$, the wave energy transforms as $\delta E \to \delta E + V \, \delta P$. The energy is not invariant under boosts, but it can be written as $\delta E = \delta E^{(0)} + \bar{u}_0^\Omega \, \delta P$, where $\delta E^{(0)}$ is the invariant energy in the ERF. Assuming the plane wave from Eq.~(\ref{eq:plane_wave}), the average wave energy per unit length in the ERF is:
\begin{eqnarray}
  \delta E^{(0)} &=& \frac{\rho \, W}{2} \left[ h_{0} \left(\delta\bar{u}^\Omega\right)^{2} + \frac{c_\Omega^{2}}{h_{0}} \left(\delta h\right)^{2}\right] \nonumber \\
  &=& \frac{\rho \, W}{4} \left[h_{0} \, k^{2} + \frac{h_{0}}{c_\Omega^{2}} \, \omega^{(0)\,2} \right] A^{2} \nonumber \\
  &=& \frac{\rho \, W}{2} \, \frac{h_{0}}{c_\Omega^{2}} \, \omega^{(0)\,2} \, A^{2} \nonumber \\
  &=& \frac{\rho \, W}{2} \, \frac{h_{0}}{c_\Omega^{2}} \, \left(\omega-\bar{u}_0^\Omega k\right)^{2} \, A^{2} \,.
\end{eqnarray}
In the laboratory frame with background flow velocity $\bar{u}_0^\Omega$, this becomes:
\begin{eqnarray}
  \delta E &=& \delta E^{(0)} + \bar{u}_0^\Omega \, \delta P \nonumber \\
  &=& \frac{\rho \, W}{2} \, \frac{h_{0}}{c_\Omega^{2}} \, \left[ \left(\omega-\bar{u}_0^\Omega k\right) + \bar{u}_0^\Omega\,k \right] \left(\omega-\bar{u}_0^\Omega k\right) \, A^{2} \nonumber \\
  &=& \frac{\rho \, W}{2} \, \frac{h_{0}}{c_\Omega^{2}} \, \omega \, \left(\omega-\bar{u}_0^\Omega k\right) \, A^{2} \,.
  \label{eq:energy}
\end{eqnarray}
Note that $\delta E/\delta P = \omega/k$, which is the phase velocity of the wave. 

A critical physical consequence of this boost transformation is that the wave energy $\delta E$ can become negative~\cite{Stepanyants-Fabrikant-1989}. This occurs whenever $\delta E^{(0)} + \bar{u}_0^\Omega \, \delta P < 0$, meaning the background flow $\bar{u}_0^\Omega$ is sufficiently fast to change the sign of the wave's phase velocity in the laboratory frame. Specifically, this occurs for counter-propagating waves whenever $|\bar{u}_0^\Omega| > c_\Omega$ (a supercritical flow). This is the fundamental hydrodynamic mechanism driving Analogue Gravity; the existence of negative-energy excitations in the interior region of the analogue black hole is what makes the Hawking process energetically possible~\cite{1975CMaPh43199H}. The total energy of the physical fluid system is strictly positive; it is only the wave energy, defined as a differential perturbation atop the moving background, that becomes negative.

\subsection{Norm and wave action with vorticity-dependent parameters}\label{subsec:norm}

We may write the momentum and energy densities in the form~\cite{Stepanyants-Fabrikant-1989}:
\begin{equation}
    \delta P = k \, \delta N \,, \qquad \delta E = \omega \, \delta N \,,
    \label{eq:P_E_N}
\end{equation}
where we have defined the norm $\delta N$:
\begin{equation}
    \delta N = \frac{\rho \, W}{2} \, \frac{h_{0}}{c_\Omega^{2}} \, \left(\omega-\bar{u}_0^\Omega k\right) \, A^{2} \,.
\end{equation}
By analogy with quantum mechanics, where excitations are quantized with momentum and energy proportional to $k$ and $\omega$, we see that $\delta N$ plays the role of the ``number of quanta.'' This is a conserved quantity. In field theory, it is referred to as the norm; in fluid mechanics, it corresponds to the {\it wave action}. In a stationary background, where the laboratory frequency $\omega$ is conserved, the conservation of norm is strictly equivalent to the conservation of wave energy. 

For a stationary mode $e^{-i\omega t}$, conservation of norm dictates the constancy of the norm flux, $\delta N \times (\omega/k)$ (in the non-dispersive limit). Recognizing that $h_{0}/c_\Omega^{2} \propto \Lambda_\Omega/c_\Omega$, and that $\rho$ and $W$ are constant, we find:
\begin{equation}
    \partial_{x}\left[ \frac{\Lambda_\Omega}{c_\Omega} \, \left(\omega-\bar{u}_0^\Omega k\right) \, \frac{\omega}{k} \, A^{2} \right] = \partial_{x}\left[ \pm \omega \, \Lambda_\Omega \, A^{2} \right] = 0 \,,
    \label{eq:norm_conservation_1}
\end{equation}
where $+$($-$) refers to the co-propagating (counter-propagating) wave. These modes always possess opposite norm fluxes, regardless of whether the flow is subcritical or supercritical. For both waves, the spatial evolution of the amplitude is governed by:
\begin{equation}
    \partial_{x}\left[ \Lambda_\Omega \, A^{2} \right] = 0 \,.
    \label{eq:norm_conservation_2}
\end{equation}
Thus, in a stationary inhomogeneous background without active mode-scattering, the wave amplitude varies precisely as $1/\sqrt{\Lambda_\Omega}$ to conserve wave action.

When wave scattering is taken into account, the amplitudes will change as modes couple. However, the {\it total} norm, summed over all products of the scattering process, remains rigorously conserved. Because the co- and counter-propagating waves carry opposite norm flux, Eq.~(\ref{eq:norm_conservation_2}) generalizes to:
\begin{equation}
    \partial_{x}\left[\Lambda_\Omega\left(A_{+}^{2}-A_{-}^{2}\right)\right] = 0 \,.
\end{equation}
By absorbing the $1/\sqrt{\Lambda_\Omega}$ factor into newly defined generic normalized amplitudes $A_{\pm} = \mathcal{A}_{\pm}/\sqrt{\Lambda_\Omega}$, and complexifying the field to capture the phase, this exact conservation law becomes:
\begin{equation}
    \partial_{x}\left[\left|\mathcal{A}_{+}\right|^{2}-\left|\mathcal{A}_{-}\right|^{2}\right] = 0 \,.
\end{equation}

Importantly, the norm $\delta N$ has no fixed sign; its sign is determined entirely by the effective co-moving frequency $\omega^{(0)} = \omega-\bar{u}_0^\Omega k$. As established in Eq.~(\ref{eq:energy}), the wave energy becomes negative precisely when $\omega^{(0)}$ and the laboratory frequency $\omega$ possess opposite signs. Because a transcritical black-hole flow forces the counter-propagating wave to flip its co-moving frequency sign as it crosses the horizon, continuous scattering in this region inherently mixes modes of opposite norm (opposite energy). 

Scattering is formally captured by the scattering matrix $S$, which dictates the linear transformation between ingoing and outgoing wave amplitudes. The strict conservation of total norm implies the unitarity relation:
\begin{equation}
    \sum_{{\rm out},{\rm norm}>0} \left| \mathcal{A}^{\rm out}_{{\rm norm}>0} \right|^{2} - \sum_{{\rm out},{\rm norm}<0} \left| \mathcal{A}^{\rm out}_{{\rm norm}<0} \right|^{2} = 1 \,.
    \label{eq:unitarity}
\end{equation}
If all outgoing waves possess positive norm, the scattering is {\it normal}, and the incident energy is simply redistributed. However, because our transcritical flow generates a negative-norm outgoing wave inside the horizon, we enter a regime of {\it anomalous} scattering. The total energy of the outgoing positive-norm waves must exceed the incident wave energy to mathematically balance the negative-norm contribution. This is the hallmark of extracting energy from the background flow.

\section{The Analogue Hawking Effect}\label{app:hawking}

As demonstrated in the main text, the transition from a subcritical region to a supercritical one fundamentally alters the nature of the counter-propagating mode. On the subcritical side, the wave successfully propagates upstream against the current, possessing positive energy. On the supercritical side, the flow overcomes the intrinsic wave speed, sweeping the wave downstream even as it attempts to propagate upstream. As discussed in Appendix~\ref{app:wave_properties}, this supercritical mode carries negative energy. Therefore, the black-hole horizon effectively splits the counter-propagating mode into two disconnected branches with opposite energies, both propagating away from the horizon. Energetically, pairs of such modes can be continuously emitted from the horizon at no overall energy cost. 

The Analogue Gravity program stems from Unruh's realization~\cite{PhysRevLett.46.1351} that this emission of opposite-energy pairs is exactly analogous to particle creation in gravitational black holes~\cite{1975CMaPh43199H}. 

\begin{figure}[h!]
    \centering
    \includegraphics[width=0.45\columnwidth]{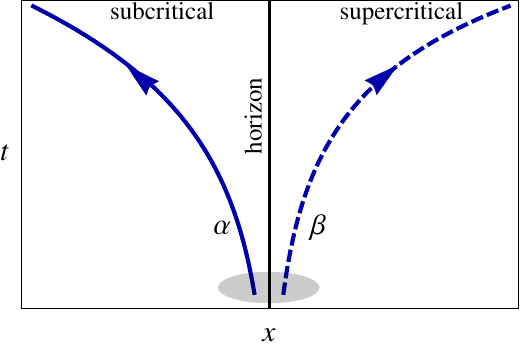} \, \includegraphics[width=0.45\columnwidth]{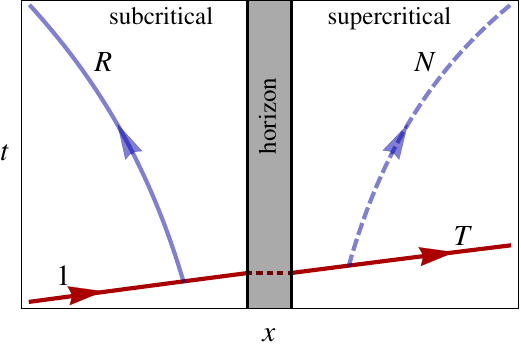}
    \caption{Space-time diagrams of two scattering processes occurring in a black-hole spacetime. The flow is from left to right, with the horizon occurring at the center. Red (blue) lines correspond to the co- (counter-) propagating mode. The negative-energy mode is shown as a dashed line. (Left) The Hawking effect: Governed by the bifurcation of the counter-propagating mode at the horizon, the relative amplitude between the two branches follows an exponential law characteristic of a Boltzmann factor. (Right) Continuous scattering over the conformal factor: The co-propagating mode crosses the horizon smoothly, but spatial variations in the effective metric induce continuous scattering into counter-propagating modes on both sides of the horizon.}
    \label{fig:scattering_processes}
\end{figure}

The simplest derivation of this phenomenon relies on the phase singularity at the horizon. As a counter-propagating wave approaches the horizon, its wave number diverges. In the near-horizon region, we can expand the characteristic velocity linearly: $|\bar{u}_0^\Omega(x)|-c_\Omega(x) \approx \kappa_\Omega \left(x-x_{\rm hor}\right)$, where $\kappa_\Omega$ is the analogue surface gravity. Applying the dispersion relation~(\ref{eq:disp_rel}) gives the wave number $k = \omega/\left(|\bar{u}_0^\Omega|-c_\Omega\right)$. The resulting waveform is:
\begin{eqnarray}
    \phi(x,t) = {\rm exp}\left(i \int^{x} k(x^{\prime})\, dx^{\prime} - i \omega t\right) &=& {\rm exp}\left(i \int^{x} \frac{\omega}{|\bar{u}_0^\Omega(x^{\prime})|-c_\Omega(x^{\prime})} \, dx^{\prime} - i \omega t\right) \nonumber \\
    &\approx& {\rm exp}\left(i \int^{x} \frac{\omega}{\kappa_\Omega\left(x^{\prime}-x_{\rm hor}\right)} \, dx^{\prime} - i \omega t\right) \nonumber \\
    &\approx& {\rm exp}\left(i \frac{\omega}{\kappa_\Omega} \, {\rm log}\left(x-x_{\rm hor}\right) \, -i\omega t\right)
\end{eqnarray}
This logarithmic phase singularity indicates that the wave's rays can be traced backward in time toward the horizon {\it ad infinitum} (as illustrated in Figure~\ref{fig:scattering_processes}). To make the field $\phi(x,t)$ physically well-behaved across this singularity, it must be analytically continued into the complex plane of $x$. 

Because the logarithm possesses a branch cut ending at $x=x_{\rm hor}$, this smooth continuation can occur entirely in either the upper- or lower-half complex plane (where $x$ acquires a positive or negative imaginary part), but not both simultaneously. This yields two independent, well-behaved modes. During the continuation of ${\rm log}\left(x-x_{\rm hor}\right)$, the term $x-x_{\rm hor}$ picks up a complex phase $e^{i\theta}$, with $\theta$ varying from $0$ to $\pm \pi$ depending on the chosen half-plane. This immediately dictates the relative amplitude between the modes localized on either side of the horizon, yielding a factor of ${\rm exp}\left(\mp \pi \omega/\kappa_\Omega\right)$. 

Letting the amplitudes of these modes be $\alpha$ and $\beta$ (with $\left|\beta\right| < \left|\alpha\right|$), their ratio is precisely $\left|\beta/\alpha\right| = {\rm exp}\left(-\pi\omega/\kappa_\Omega\right)$. Normalizing these amplitudes to satisfy the unitarity relation $\left|\alpha\right|^{2}-\left|\beta\right|^{2} = 1$ (the minus sign accounting for the opposite energy of the modes, per Eq.~\ref{eq:unitarity}), we find:
\begin{equation}
    \left|\beta\right|^{2} = \frac{1}{e^{2\pi \omega/\kappa_\Omega}-1} \,, \qquad {\rm where} \qquad \kappa_\Omega = \left.\partial_{x}\left(|\bar{u}_0^\Omega|-c_\Omega\right)\right|_{\rm hor} = \frac{1}{2|\bar{u}_0^\Omega|} \left. \partial_{x}\left((\bar{u}_0^\Omega)^{2}-c_\Omega^{2}\right) \right|_{\rm hor} \,.
    \label{eq:Hawking-Unruh}
\end{equation}
This quantifies the exact rate of ``pair creation''—the stimulated emission of opposite-energy waves on either side of the horizon—required for the wavefield to remain mathematically continuous. Strikingly, this takes the exact form of a thermal Planck spectrum, allowing us to read off the analogue Hawking temperature as $k_{B}T_H^\Omega = \hbar\kappa_\Omega/2\pi$. While the presence of $\hbar$ highlights its quantum mechanical origins, the emission is fundamentally governed by the classical characteristic frequency $\kappa_\Omega/2\pi$, making this framework perfectly applicable to stimulated emission in macroscopic fluid systems.

\section{Field theory formulation\label{app:field_theory}}

In this appendix, we briefly introduce the effective field theory description of the system that would typically be used in the context of curved spacetime.  The point is simply to show that such a mapping can be made.  We also emphasize that the formalism used here brings with it a powerful mathematical structure.

\subsection{Lagrangian and canonical fields}

The action associated with a massless scalar field has the form~\cite{peskin2018introduction,Birrell_Davies_1982}
\begin{equation}
    S = \iint dt \, dx \, \sqrt{-g} \, \mathcal{L} \,, \qquad \mathcal{L} = -\frac{1}{2} \, g^{\mu\nu} \, \partial_{\mu}\phi \, \partial_{\nu}\phi \,.
\end{equation}
$\mathcal{L}$ is itself a scalar, while $\sqrt{-g} \, dx \, dt$ is the invariant volume element of the spacetime (that is, both are invariant under general coordinate transformations).  The canonical field $\phi$ comes with an associated canonical momentum $\pi$:
\begin{equation}
    \pi = \frac{\partial\left(\sqrt{-g} \, \mathcal{L}\right)}{\partial\left(\partial_{t}\phi\right)} = -\sqrt{-g} \, g^{t \, \mu} \partial_{\mu}\phi = -\sqrt{-g}\left(g^{tt}\partial_{t}+g^{tx}\partial_{x}\right)\phi \,.
\end{equation}
Given the general metric in Eq.~\eqref{eq:intro_metric}, and that for simplicity we report below
\begin{equation}
    \left[\left(\partial_t+\partial_xU\right)\frac{\Lambda}{c}\left(\partial_t+U\partial_x\right)-\partial_x\Lambda c\partial_x\right]\phi=0\,,
\end{equation}
we have
\begin{equation}
    g_{\mu\nu} = \Lambda^{2} \left[ \begin{array}{ccc} -\left(c^{2}-U^{2}\right) & -U & 0 \\ -U & 1 & 0 \\ 0 & 0 & 1 \end{array} \right] \,, \qquad g^{\mu\nu} = \frac{1}{\Lambda^{2}c^{2}} \left[ \begin{array}{ccc} -1 & -U & 0 \\ -U & c^{2}-U^{2} & 0 \\ 0 & 0 & c^{2} \end{array} \right] \,, \qquad \sqrt{-g} = \Lambda^{3}c \,.
\end{equation}
Thus, considering a flow characterized by a constant shear, the effective metric which describes the effective spacetime is the one in Eq.~\eqref{eq:wave_eqn_on_shear_flow}. Then, 
our canonical field $\phi$ is just the perturbation of the velocity potential $\delta\varphi^\Omega$, while the metric parameters are fixed as
\begin{equation}
    \label{eq:parameters3}
    U=\bar{u}_0^\Omega=u^{(s)}_0-\frac{\Omega h_0}{2}\,,\qquad c^2=c^2_\Omega=gh_0+\frac{\Omega^2h_0^2}{4}\,,\qquad\Lambda^2=\Lambda_\Omega^2=\frac{h_0/h_{\rm up}}{1+\frac{\Omega^2h_0}{4g}}\,.
\end{equation}
Finally, the
canonical momentum is
\begin{equation}
    \pi = \frac{\Lambda}{c} \left(\partial_{t}+U\partial_{x}\right)\phi\qquad\Longrightarrow\qquad \pi^\Omega=\frac{\Lambda_\Omega}{c_\Omega} \left(\partial_{t}+\bar{u}_0^\Omega\partial_{x}\right)\delta\varphi^\Omega=\frac{1}{1+\frac{\Omega^{2}h_{0}}{4g}}\left(\partial_{t}+\left(u^{(s)}_0-\frac{\Omega h_0}{2}\right)\partial_{x}\right)\delta\varphi^\Omega = g \, \delta h \,.
\end{equation}

\subsection{Hamiltonian}

Given a general canonical field $\phi$, with conjugated momentum $\pi$, propagating on top of a curved spacetime described by the metric in Eq.~\eqref{eq:intro_metric}, the energy density is identified with the Hamiltonian density, which is given by
\begin{eqnarray}
    \mathcal{H} &=& \pi \, \partial_{t}\phi - \sqrt{-g} \, \mathcal{L} \nonumber \\
    &=& \frac{\Lambda }{c }\left(\partial_{t}+U\partial_{x}\right)\phi \, \partial_{t}\phi - \frac{\Lambda }{c } \left\{ \frac{1}{2} \left[\left(\partial_{t}+U\partial_{x}\right)\phi\right]^{2} - \frac{c ^{2}}{2} \left(\partial_{x}\phi\right)^{2} \right\} \nonumber \\
    &=& \frac{\Lambda }{c } \left\{ \frac{1}{2}\left[\left(\partial_{t}+U\partial_{x}\right)\phi\right]^{2} + \frac{c ^{2}}{2}\left(\partial_{x}\phi\right)^{2} - U \, \partial_{x}\phi \, \left(\partial_{t}+U\partial_{x}\right)\phi \right\} \,.
\end{eqnarray}
Considering again a flow characterized by a constant shear, the effective metric which describes the effective spacetime is characterized by the parameters in Eq.~\eqref{eq:parameters3}. Then, the Hamiltonian $\mathcal{H}^\Omega$ 
 agrees (up to a constant prefactor) with the energy density of Appendix~\ref{sec:Wave_energy}, and we clearly see the appearance of the cross-term that allows the energy to be negative.  In fact this can be rewritten in an even simpler form:
\begin{equation}
    \mathcal{H}^\Omega = \frac{\Lambda_\Omega}{c_\Omega}\left\{ \frac{1}{2} \left(\partial_{t}\delta\varphi^\Omega\right)^{2} + \frac{c_\Omega^{2}-(\bar{u}_0^\Omega)^{2}}{2} \left(\partial_{x}\delta\varphi^\Omega\right)^{2} \right\} \,,
\end{equation}
where we see clearly that the energy is allowed to be negative only when $(\bar{u}_0^\Omega)^2 > c_\Omega^2$ ({\it i.e.}, when ${\rm Fr_\Omega}>1$).

\subsection{Scalar product and norm}

We finish this section by noting an important symmetry and its associated conservation law.  Let us complexify the field $\phi$, whereupon the Lagrangian becomes
\begin{equation}
    \mathcal{L} = -g^{\mu\nu}\partial_{\mu}\phi^{\star} \partial_{\nu}\phi \,. 
\end{equation}
Although the actual fields we consider are real, there is nothing to prevent us from generalizing the field to a complex one when the wave equation we work with is linear.  The key advantage is that it allows us to exploit the symmetry of the Lagrangian under phase rotations of the field, $\phi \to \phi \, e^{i\theta}$.  This engenders, via Noether's theorem, an associated conservation law, which is:
\begin{equation}
    \partial_{t}\left(\phi_{1}\,,\,\phi_{2}\right) = 0 \,, \qquad \left(\phi_{1}\,,\,\phi_{2}\right) = i\int dx \, \frac{\Lambda}{c} \Big[ \phi_{1}^{\star} \, \left(\partial_{t}+U\partial_{x}\right)\phi_{2} - \phi_{2} \, \left(\partial_{t}+U\partial_{x}\right)\phi_{1}^{\star} \Big] \,.
\end{equation}
The structure $\left(\phi_{1}\,,\,\phi_{2}\right)$ defines a scalar product on the space of all wave solutions.
The scalar product of a solution with itself is often called its {\it norm}, and one can easily show that it coincides (up to a constant factor) with the norm defined in Appendix~\ref{subsec:norm}.  We see clearly that, although it is a real number, it is not necessarily positive: indeed, if a solution $\phi$ has positive norm, then $\phi^{\star}$ (which is also a solution because the wave equation has real coefficients) has negative norm.  We also see that real solutions have zero norm. 

\section{Implications of the conformal factor}\label{app:conformal_factor}

In this appendix, we detail the mathematical consequences of a spatially varying conformal factor $\Lambda(x)$ within the generalized Painlevé-Gullstrand metric. We first derive the coupled amplitude equations governing continuous mode-mixing (scattering), and subsequently demonstrate how this spatial variation manifests exactly as an effective Regge-Wheeler-like potential barrier when the wave equation is cast into a Schrödinger-like form.

\subsection{Derivation of mode mixing\label{subsec:mode_mixing}}

Let us consider a generic effective spacetime described by the metric in Eq.~\eqref{eq:intro_metric}.
The corresponding equation of motion for a field $\phi$ propagating on this metric is:
\begin{equation}
    \label{eq:wave_eqn_general}
    \Box\phi=\left[\left(\partial_t+\partial_xU\right)\frac{\Lambda}{c}\left(\partial_t+U\partial_x\right)-\partial_x\Lambda c\partial_x\right]\phi=0\,.
\end{equation}
We introduce two plane waves:
\begin{equation}
    w_{\pm}(x) = {\rm exp}\left[i \int^{x} k_{\pm}\left(x^{\prime}\right) \, dx^{\prime} \right] \,, \qquad k_{\pm}\left(x^{\prime}\right) = \frac{\omega}{U\left(x^{\prime}\right) \pm c\left(x^{\prime}\right)} \,.
\end{equation}
These are chosen because they are exact solutions to the wave equation whenever the generic conformal factor $\Lambda$ is strictly constant. To see the extent to which they fail to solve the full wave equation when $\Lambda$ varies, we plug them into Eq.~\eqref{eq:wave_eqn_general}. First, note that:
\begin{equation}
    \partial_{x}w_{\pm} = i k_{\pm}\, w_{\pm} \,, \qquad \partial_{x}^{2}w_{\pm} = \left(-k_{\pm}^{2} + i k_{\pm}^{\prime}\right) w_{\pm} \,.
\end{equation}
By setting $\partial_{t} \to -i\omega$ and expanding, the d'Alembertian of Eq.~\eqref{eq:intro_metric} becomes:
\begin{equation}
   \Box = -\frac{\Lambda}{c}\omega^{2} -i\omega\left(U \frac{\Lambda}{c}\right)^{\prime} -2 i \omega U \frac{\Lambda}{c} \partial_{x} + \left[\frac{\Lambda}{c}\left(U^{2}-c^{2}\right)\right]^{\prime} \partial_{x} + \frac{\Lambda}{c}\left(U^{2}-c^{2}\right) \partial_{x}^{2} \,.
   \label{eq:dAlembertian_fixed_omega}
\end{equation}
Applying this to the plane waves $w_{\pm}$ yields:
\begin{eqnarray}
    \Box w_{\pm} &=& \left[ -\frac{\Lambda}{c}\omega^{2} -i\omega\left(U \frac{\Lambda}{c}\right)^{\prime} + 2 \omega U \frac{\Lambda}{c} k_{\pm} + i \left[\frac{\Lambda}{c}\left(U^{2}-c^{2}\right)\right]^{\prime} k_{\pm} - \frac{\Lambda}{c}\left(U^{2}-c^{2}\right) k_{\pm}^{2} + i \frac{\Lambda}{c}\left(U^{2}-c^{2}\right) k_{\pm}^{\prime} \right] w_{\pm} \nonumber \\
    &=& \left[ -\frac{\Lambda}{c} \left\{ \left(\omega-Uk_{\pm}\right)^{2} - c^{2} k_{\pm}^{2} \right\} + i \left[\frac{\Lambda}{c}\left\{ \left(U^{2}-c^{2}\right) k_{\pm} - \omega U \right\} \right]^{\prime} \right] w_{\pm} \nonumber \\
    &=& i \left[ \frac{\Lambda}{c} \left\{ \mp \omega c \right\} \right]^{\prime} w_{\pm} \nonumber \\
    &=& \mp i \omega \, \Lambda^{\prime} \, w_{\pm} \,.
    \label{eq:Box-w}
\end{eqnarray}
This confirms that $w_{\pm}$ only solve the wave equation if $\Lambda^{\prime} = 0$.

To combine $w_{\pm}(x)$ into an exact solution of the full wave equation, we introduce position-dependent amplitudes $a_{\pm}(x)$, such that:
\begin{equation}
    \phi = a_{+} w_{+} + a_{-} w_{-} \,, \qquad \phi^{\prime} = a_{+} w_{+}^{\prime} + a_{-} w_{-}^{\prime} \,.
    \label{eq:phi_and_first_derivative}
\end{equation}
The first equation decomposes the exact solution into the plane waves, while the second acts as a constraint, imposing that the derivative satisfies exactly the same decomposition. This constraint greatly simplifies the calculations and is mathematically equivalent to requiring:
\begin{equation}
    a_{+}^{\prime} w_{+} + a_{-}^{\prime} w_{-} = 0 \,.
    \label{eq:constraint}
\end{equation}
Furthermore, it yields the following expression for the second derivative of $\phi$:
\begin{equation}
    \phi^{\prime\prime} = a_{+} w_{+}^{\prime\prime} + a_{-} w_{-}^{\prime\prime} + a_{+}^{\prime} w_{+}^{\prime} + a_{-}^{\prime} w_{-}^{\prime} \,.
    \label{eq:phi_second_derivative}
\end{equation}
Examining Eqs.~(\ref{eq:phi_and_first_derivative}) and~(\ref{eq:phi_second_derivative}), we see that $\phi^{(n)} = a_{+}w_{+}^{(n)} + a_{-}w_{-}^{(n)}$, with an additional contribution appearing only in the second derivative. Applying the d'Alembertian~(\ref{eq:dAlembertian_fixed_omega}) to $\phi$, we obtain:
\begin{equation}
    \Box \phi = a_{+} \Box w_{+} + a_{-} \Box w_{-} + \frac{\Lambda}{c} \left(U^{2}-c^{2}\right) \left( a_{+}^{\prime} w_{+}^{\prime} + a_{-}^{\prime} w_{-}^{\prime} \right) = 0 \,.
\end{equation}
Substituting the results from Eq.~(\ref{eq:Box-w}) gives:
\begin{eqnarray}
  \Box \phi &=& -i \omega \, \Lambda^{\prime} \left( a_{+} w_{+} - a_{-} w_{-} \right) + \frac{\Lambda}{c} \left(U^{2}-c^{2}\right) \left( a_{+}^{\prime} w_{+}^{\prime} + a_{-}^{\prime} w_{-}^{\prime} \right) \nonumber \\
  &=& -i \omega \left[ \Lambda^{\prime} \left( a_{+} w_{+} - a_{-} w_{-} \right) - \frac{\Lambda}{c}\left(U^{2}-c^{2}\right) \left(\frac{1}{U+c} \, a_{+}^{\prime} w_{+} + \frac{1}{U-c} \, a_{-}^{\prime} w_{-} \right) \right] \nonumber \\
  &=& -i \omega \left[ \Lambda^{\prime} \left( a_{+} w_{+} - a_{-} w_{-} \right) - \frac{\Lambda}{c}\left(\left(U-c\right) a_{+}^{\prime} w_{+} + \left(U+c\right) a_{-}^{\prime} w_{-} \right) \right] \nonumber \\
  &=& -i \omega \left[ \Lambda^{\prime} \left( a_{+} w_{+} - a_{-} w_{-} \right) + \Lambda \left( a_{+}^{\prime} w_{+} - a_{-}^{\prime} w_{-} \right) \right] = 0 \,.
\end{eqnarray}
Recalling the constraint in Eq.~(\ref{eq:constraint}), we isolate $a_{\pm}^{\prime}$:
\begin{equation}
    a_{+}^{\prime} = -\frac{\Lambda^{\prime}}{2\Lambda} \left(a_{+}-a_{-}\frac{w_{-}}{w_{+}}\right) \,, \qquad a_{-}^{\prime} = - \frac{\Lambda^{\prime}}{2\Lambda} \left(a_{-}-a_{+}\frac{w_{+}}{w_{-}}\right) \,.
\end{equation}
Finally, we eliminate the diagonal contribution to the evolution of $a_{\pm}$ by defining the generic normalized amplitudes $\mathcal{A}_{\pm}$:
\begin{equation}
    a_{\pm} = \frac{\mathcal{A}_{\pm}}{\sqrt{\Lambda}} \,.
\end{equation}
The evolution equations then simplify to:
\begin{equation}
    \mathcal{A}_{+}^{\prime} = \frac{\Lambda^{\prime}}{2\Lambda} \, \frac{w_{-}}{w_{+}} \, \mathcal{A}_{-} \,, \qquad \mathcal{A}_{-}^{\prime} = \frac{\Lambda^{\prime}}{2\Lambda} \, \frac{w_{+}}{w_{-}} \, \mathcal{A}_{+} \,,
\end{equation}
or, expressed in matrix form:
\begin{equation}
    \label{eq:amplitudes-general}
    \partial_{x}\left[\begin{array}{c} \mathcal{A}_{+} \\ \mathcal{A}_{-} \end{array} \right] = \frac{\Lambda^{\prime}}{2\Lambda} \left[ \begin{array}{cc} 0 & {\rm exp}\left(2i \omega \int^{x} \frac{c}{U^{2}-c^{2}} \, dx^{\prime}\right) \\ {\rm exp}\left(-2i \omega \int^{x} \frac{c}{U^{2}-c^{2}} \, dx^{\prime}\right) & 0 \end{array} \right] \, \left[ \begin{array}{c} \mathcal{A}_{+} \\ \mathcal{A}_{-} \end{array} \right] \,.
\end{equation}
This establishes a direct coupling between the two mode amplitudes. A mathematically analogous result governs pair creation in cosmology~\cite{Zel'dovich1979,Massar-Parentani-1998}.

\subsection{Effective scattering potential}\label{subsec:effectivepotential}

As discussed in Sec.~\ref{sec:Wave_scattering}, the spatial variation of the conformal factor $\Lambda(x)$ in the generic metric (Eq.~\eqref{eq:intro_metric}) induces mode-mixing. This continuous scattering can be interpreted as the effect of a potential barrier acting on the modes, in perfect analogy with the Regge-Wheeler potential in General Relativity~\cite{PhysRev.108.1063,Anderson2014}. 

Here, we derive the exact form of this effective potential. We consider a massless scalar field $\phi$ propagating on the generic effective spacetime described by Eq.~\eqref{eq:intro_metric}. Its equation of motion is:
\begin{equation}
    \left[\left(\partial_t+\partial_xU\right)\frac{\Lambda}{c}\left(\partial_t+U\partial_x\right)-\partial_xc\Lambda\partial_x\right]\phi=0\,.
\end{equation}
To extract the effective potential, we perform a series of coordinate transformations. First, we define the \emph{Schwarzschild time} $t_s$:
\begin{equation}
    dt_s=dt+\frac{U}{c^2-U^2}dx\,.
\end{equation}
Consequently, the wave equation takes a \emph{Schwarzschild form}:
\begin{equation}
    \left[\frac{\Lambda/c}{1-\frac{U^2}{c^2}}\partial_{t_s}^2-\partial_x\frac{\Lambda}{c}(c^2-U^2)\partial_x\right]\phi=0\,.
\end{equation}
Second, we reparameterize the spatial coordinate by introducing the \emph{tortoise coordinate} $\tau_*$, defined as:
\begin{equation}
    \label{eq:tortoise-coordinate}
    d\tau_*=\frac{dx/c}{1-\frac{U^2}{c^2}}\,.
\end{equation}
Note that the differential $d\tau_*$ diverges at the analogue horizon ($U=c$). On either side of the horizon, the tortoise coordinate diverges with opposite signs: approaching the horizon from the subcritical region yields $\tau_* \to +\infty$, while approaching from the supercritical region yields $\tau_* \to -\infty$. In terms of the tortoise coordinate, the wave equation simplifies to:
\begin{equation}
    \left(\partial_{t_s}^2-\frac1\Lambda\partial_{\tau_*}\Lambda\partial_{\tau_*}\right)\phi=0\,.
\end{equation}
By introducing the rescaled field $u = \phi\sqrt{\Lambda}$ and restricting our analysis to stationary modes of the form $u(t_s,\tau_*) = e^{-i\omega t_s} u_\omega(\tau_*)$, the equation can be recast as an \emph{effective Schrödinger equation}:
\begin{equation}
    \left(\omega^2+\partial_{\tau_*}^2-V_{\rm eff}(\tau_*)\right)u_\omega=0\,,\qquad V_{\rm eff}(\tau_*)=\frac{1}{\sqrt{\Lambda}}\frac{d^2}{d\tau_*^2}\sqrt{\Lambda}\,.
\end{equation}
The final step is to express this effective potential $V_{\rm eff}$ in terms of the original spatial coordinate $x$. Using the definition of the tortoise coordinate in Eq.~\eqref{eq:tortoise-coordinate}, the spatial derivatives transform as:
\begin{align}
    \partial_{\tau_*}&=\frac{\partial x}{\partial\tau_*}\partial_x=c\left(1-\frac{U^2}{c^2}\right)\partial_x\,,\\
    \partial_{\tau_*}^2&=c\left(1-\frac{U^2}{c^2}\right)\partial_x \left[ c\left(1-\frac{U^2}{c^2}\right)\partial_x \right] \nonumber\\
    &=c^2\left(1-\frac{U^2}{c^2}\right)^2\partial_x^2+\left(1-\frac{U^2}{c^2}\right)\left[\left(1+\frac{U^2}{c^2}\right)c\,c'-2U\,U'\right]\partial_x\,,
\end{align}
where primes denote derivatives with respect to $x$. Substituting these expressions into the definition of the effective potential, we obtain the exact sought-after function:
\begin{equation}
    \label{eq:effectivepotential}
    V_{\rm eff}=c^2\left(1-\frac{U^2}{c^2}\right)^2\left(\frac{\Lambda''}{2\Lambda}-\frac{(\Lambda')^2}{4\Lambda^2}\right)+\left(1-\frac{U^2}{c^2}\right)\left[\left(1+\frac{U^2}{c^2}\right)c\,c'-2U\,U'\right]\frac{\Lambda'}{2\Lambda}\,.
\end{equation}

\section{Wave scattering over an asymmetrical obstacle\label{app:asymmetrical}}

Figure~\ref{fig:scattering_amplitudes_fixed-q} was
produced assuming an obstacle with a Gaussian profile~(\ref{eq:gaussian_obstacle}).  In reality there are advantages to considering an asymmetrical profile, particularly one with a mild slope on the downstream side so as to minimize boundary separation and recirculation~\cite{Chaline2013,Weinfurtner_2011,Euve_2016,Euv__2020,Fourdrinoy_2022,CRPHYS_2024__25_G1_457_0}.

Here, we show the scattering coefficients pertaining to an asymmetrical obstacle, of similar shape to obstacles used in experiments at Institut Pprime.  The shape of this obstacle is simply a trapezoid:
\begin{equation}
    \frac{b(x)}{b_{\rm max}} =
    \begin{cases}
    \frac{x-x_{1}}{x_{2}-x_{1}} & x_{1} \le x \le x_{2} \\[0.1cm]
    1 & x_{2} \le x \le x_{3} \\[0.1cm]
    \frac{x_{4}-x}{x_{4}-x_{3}} & x_{3} \le x \le x_{4} \\[0.1cm]
    \end{cases},
    \label{eq:ACRI2010_obstacle}
\end{equation}
where $b_{\rm max} = 10\,{\rm cm}$, $x_{1} = -0.53\,{\rm m}$, $x_{2} = -0.23\,{\rm m}$, $x_{3} = 0.23\,{\rm m}$, and $x_{4} = 0.98\,{\rm m}$.

Note that this obstacle has sharp corners, which is not strictly compatible with the long-wavelength limit in which the effective metric description is valid.  This is particularly problematic for the analogue Hawking effect, which is concerned with the behavior of the flow near the horizon and is governed mainly by the second derivative of the obstacle profile at the top of the obstacle (see Eq.~\eqref{eq:surface_gravity}).  For the piecewise obstacle used here, with a flat top of finite extension, this second derivative is not well-defined.  However, as far as mixing between the co- and counter-propagating modes is concerned -- and especially for the scattering of an incident co-current mode, where the amplitudes of the counter-propagating waves is zero at the horizon and the phase singularity is thereby not encountered -- then the evolution equations derived in Sec.~\ref{sec:Wave_scattering} remain sensible and well-defined.  The initial conditions need simply be specified a short distance $\epsilon$ on either side of the flat top ({\it i.e.}, at $x=x_{2}-\epsilon$ and $x=x_{3}+\epsilon$) instead of at $x=x_{\rm hor}\pm \epsilon$ as in Sec.~\ref{sec:Wave_scattering}.

\begin{figure}
    \centering
    \includegraphics[width=0.9\columnwidth]{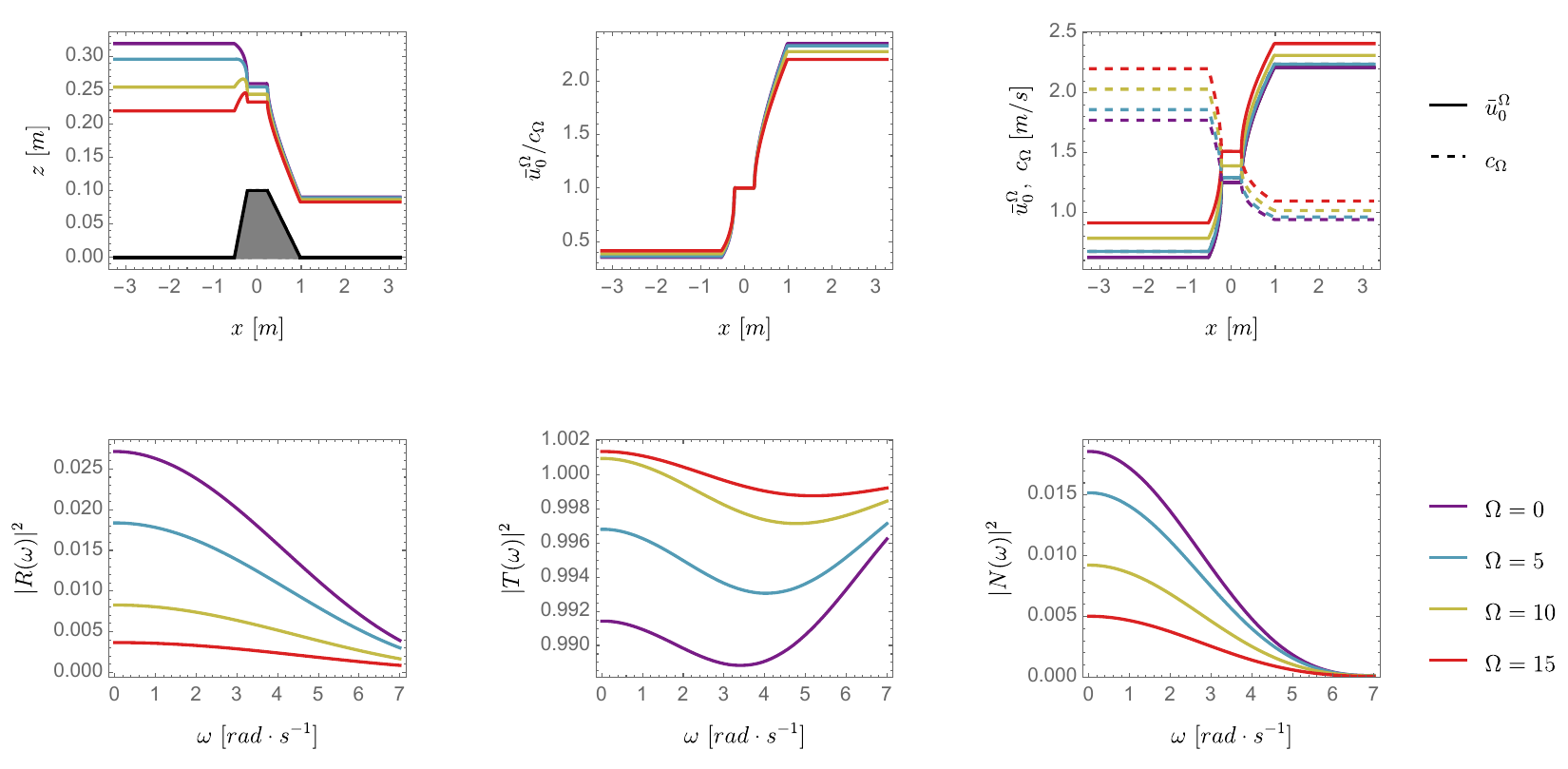}
    \caption{Scattering coefficients over a fixed asymmetric obstacle (described in Eq.~(\ref{eq:ACRI2010_obstacle})) with identical values of the flow rate $q = 0.2\,{\rm m}^{2}/{\rm s}$, across several values of transverse background vorticity (in units of $\mathrm{s}^{-1}$); this includes the irrotational flow with $\Omega=0$.  The top row shows the background flow: the water depth (left), the Froude number (center), and the profiles of $\bar{u}_{0}^\Omega$ and $c_\Omega$ (right).  On the bottom are shown the scattering coefficients for the reflected mode (left), the transmitted mode (center), and the negative-energy mode (right).  We notice that the scattering is suppressed for larger values of $\Omega$.}
    \label{fig:scattering_amplitudes_fixed-q_acri}
\end{figure}

The results are shown in 
Fig.~\ref{fig:scattering_amplitudes_fixed-q_acri}, where again the flow rate $q$ has been fixed at $0.2\,{\rm m}^{2}/{\rm s}$, exactly as in Fig.~\ref{fig:scattering_amplitudes_fixed-q}.
The sharp corners seen in the flow profile are inherited from the sharp corners of the obstacle as a result of our idealized treatment in the long-wavelength limit, and of course will be smoothed out in an actual experiment.  The results for the scattering coefficients are very similar to what we saw in Sec.~\ref{sec:Wave_scattering}, particularly the fact that the scattering tends to be suppressed with increasing background vorticity.  The main visible difference here is that the transmission coefficient does not exceed 1 as clearly as it did in 
Fig.~\ref{fig:scattering_amplitudes_fixed-q}.
This indicates a shift in the ratio between the scattering into the negative-energy mode on the downstream side of the obstacle, compared with the scattering into the positive-energy reflected mode on the upstream side.  Indeed, comparing 
Fig.~\ref{fig:scattering_amplitudes_fixed-q_acri} with Fig.~\ref{fig:scattering_amplitudes_fixed-q},
we see that $\left|R\right|^{2}$ falls off quickly at higher frequencies in 
Fig.~\ref{fig:scattering_amplitudes_fixed-q}
while $\left|N\right|^{2}$ has an extended tail, whereas in 
Fig.~\ref{fig:scattering_amplitudes_fixed-q_acri}
it is the other way round.  This makes sense when we consider that, in 
Fig.~\ref{fig:scattering_amplitudes_fixed-q_acri},
the downstream part of the flow varies more slowly by construction of the obstacle, and as the scattering amplitude is related to the derivative of $\Lambda_\Omega(x)$ then we can expect a suppression of $\left|N\right|^{2}$.

\begin{figure}
    \centering
    \includegraphics[width=0.9\columnwidth]{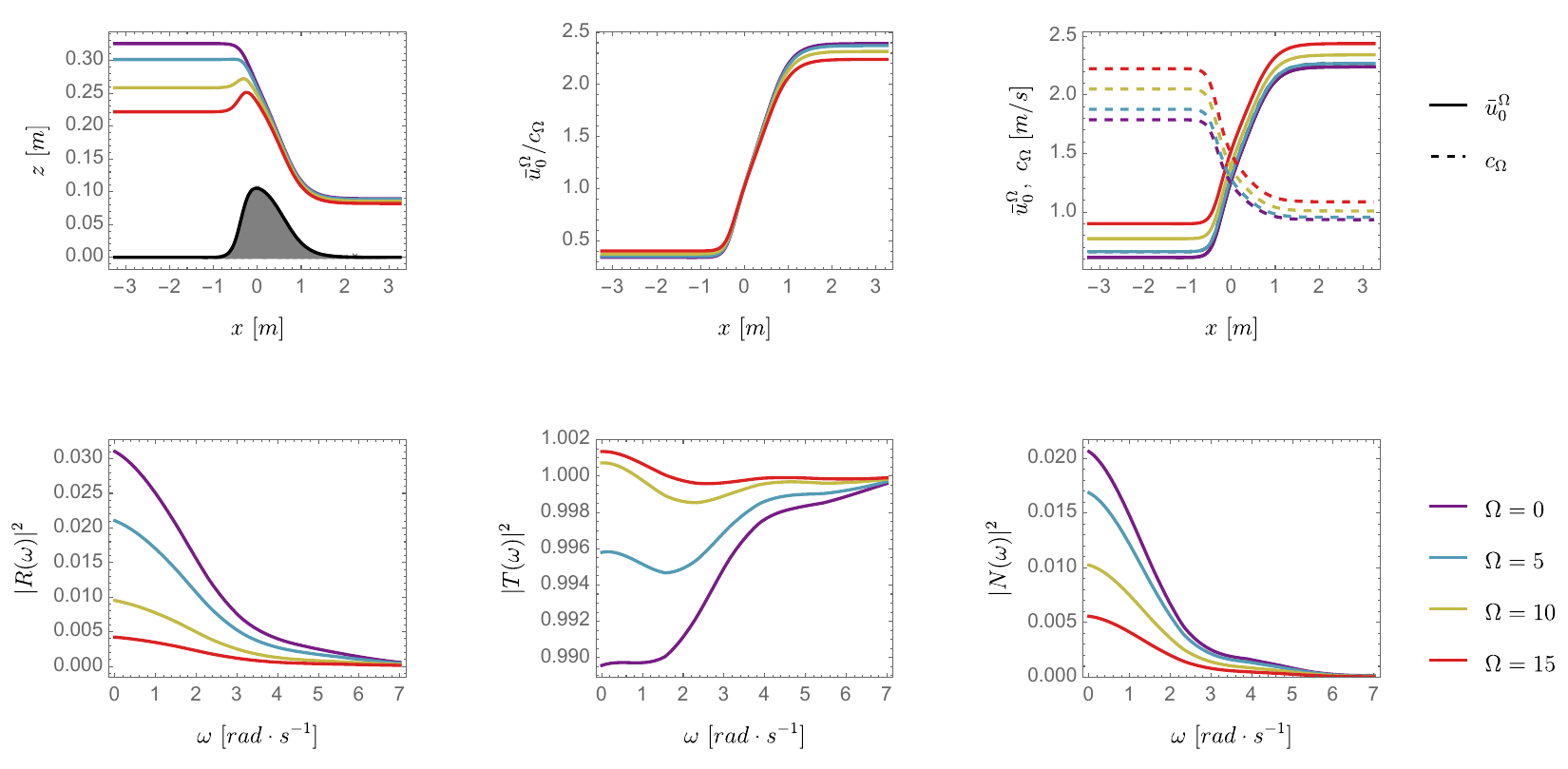}
    \caption{Scattering coefficients over a fixed asymmetric obstacle (described in Eq.~(\ref{eq:ACRI2010_obstacle_smoothed}), which is a smoothed version of that in Eq.~(\ref{eq:ACRI2010_obstacle})) with identical values of the flow rate $q = 0.2\,{\rm m}^{2}/{\rm s}$, across several values of transverse background vorticity (in units of $\mathrm{s}^{-1}$); this includes the irrotational flow with $\Omega=0$.  The top row shows the background flow: the water depth (left), the Froude number (center), and the profiles of $\bar{u}_{0}^\Omega$ and $c_\Omega$ (right).  On the bottom are shown the scattering coefficients for the reflected mode (left), the transmitted mode (center), and the negative-energy mode (right).  We notice that the scattering is suppressed for larger values of $\Omega$.}
    \label{fig:scattering_amplitudes_fixed-q_acri-smoothed}
\end{figure}

As a final check, we also consider the scattering over an obstacle which is essentially a smoothed-out version of that of Eq.~(\ref{eq:ACRI2010_obstacle}).  This obstacle has the following form:
\begin{equation}
    b(x) = b_{0} \left[ {\rm tanh}\left(m_{A}\left(x-x_{A}\right)\right) - {\rm tanh}\left(m_{B}\left(x-x_{B}\right)\right) \right] \,,
    \label{eq:ACRI2010_obstacle_smoothed}
\end{equation}
where $b_{0} = 6\,{\rm cm}$, $m_{A}=4.74\,{\rm m}^{-1}$, $x_{A}=-0.375\,{\rm m}$, $m_{B} = 1.87\,{\rm m}^{-1}$, and $x_{B} = 0.602\,{\rm m}$.  Note that $b_{\rm max}$ here is very slightly larger than before, at $10.5\,{\rm cm}$.  The associated scattering coefficients are presented in 
Fig.~\ref{fig:scattering_amplitudes_fixed-q_acri-smoothed}.
They behave in a qualitatively similar way, the main difference being that they are suppressed more quickly at larger frequencies.

\end{appendices}

\bibliography{bib}
\end{document}